\documentclass[notitlepage, 10pt]{revtex4-1}

\usepackage{amsmath}    
\usepackage{graphicx}   
\usepackage{float}
\restylefloat{table}

\begin{document}

\title{Explanation of the odd structure of fractional Hall states in higher Landau levels and filling ratios with even denominators}

\author{Janusz Jacak}
\email{janusz.jacak@pwr.edu.pl}
\affiliation{Institute of Physics, Wroc{\l}aw University of Technology, Wyb. Wyspia\'nskiego 27, 50-370 Wroc{\l}aw, Poland}
\author{Lucjan Jacak}
\email{lucjan.jacak@pwr.edu.pl}
\affiliation{Institute of Physics, Wroc{\l}aw University of Technology, Wyb. Wyspia\'nskiego 27, 50-370 Wroc{\l}aw, Poland}

\begin{abstract}
The structure of fractional fillings of higher Landau levels including spin subbands is systematically derived for the first time. Using topology-type commensurability arguments for 2D charged system in the presence of strong quantizing magnetic field, a hierarchy of  fillings in  the higher Landau levels for FQHE and for other correlated states is determined in perfect agreement with the experimental observations. The relative paucity of fractional structure in the higher Landau levels in contrast to the plethora of filling factors for FQHE in the zeroth Landau level is explained. The filling fractions with even denominators in consecutive  LLs were also identified together with the criterion for particle pairing.
\end{abstract}


\maketitle

\section{Introduction}
Significant difference in experimentally observed manifestation of FQHE in higher Landau levels (LLs) \cite{lls,ll8/3-1,5/2-1,ll8/3-2} in comparison to the lowest LL (LLL) case \cite{pan2003} has not found an explanation as of yet. In particular, the theory of composite fermions (CFs) \cite{jain,jain2007} successfully developed for the LLL, failed in systematic description of fractional hierarchy of correlated states in higher LLs. Nevertheless, the topology based theory of FQHE \cite{jac1,ws} may be applied to the lowest and to the higher LLs as well. 
This creates possibility to gain insight into structure of fillings of higher LLs when correlated multiparticle states may be arranged not along the scheme of standard CF model of which explanation ability is limited rather to the LLL. The comparison with experimental observations strongly supports the cyclotron braid group approach basing on perfect coincidence of  the theoretical prediction of specific filling ratios with the measurement data in few first LLs. The topological arguments directly explain why in the higher LLs the FQHE structure is so scanty in comparison to  the plethora of fractional Hall correlated states in the LLL. 
The idea of multilooped cyclotron orbits related to cyclotron braid subgroup seems to be confirmed, on the other hand, by experiments in suspended graphene revealing the triggering role of mean free path length proportional to carrier mobility in formation of collective FQHE state \cite{fqhe2,jac2013}, cf. also Refs \cite{bolotin2008}, \cite{pfeiffer2003} and \cite{wspolm}. Moreover, the systematic analysis of fillings of higher LLs supplies information on how to model wave functions for various fillings in terms of Halperin multicomponent generalization of the Laughlin function \cite{halperin1983,goerbig2007}.

\subsection{Microscopic models for FQHE revisited}

For fractional fillings of the lowest Landau level, $\nu=\frac{p}{q}$, $q$--odd integer and $p$--odd or even number, have been observed characteristic electric transport features referred as to FQHE \cite{pan2003}. With improving sample quality (increasing mobility of carriers) accompanying the development of measurement technique, still new ratios for FQHE were recorded (now over 70 including both spin subbands of the LLL, i.e., for $0<\nu<2$). The corresponding resistivity minima resemble Shubnikov-deHaas oscillations around fractions with even denominators,
like $\frac{1}{2}$, $\frac{1}{4}$. Such  rich manifestation of FQHE in the LLL was a puzzle how to combine it with the famous explanation of FQHE at $\nu =\frac{1}{q}$ given by Laughlin \cite{laughlin1,laughlin2}.

Enhancement or reduction of $\nu$ in vicinity of $\frac{1}{q}$ produces excess or deficiency of particles in strongly correlated Laughlin state called as quasiparticles or quasiholes with respect to the initial state. 
Halperin \cite{hh1} and Haldane \cite{hh2} proposed to reproduce the experimentally observed hierarchy of fractional fillings via concept of Laughlin condensation of these quasiparticles or quasiholes beyond or beneath the initial $\frac{1}{q}$ state. The daughter subsystem could reveal itself correlated fractional state with again own subsequent generation of quasiparticles and quasiholes resulting then in the possibility of next generation of the condensate. Filling factors corresponding to this construction \cite{hh1,hh2}, called as HH hierarchy of LL fillings, 
\begin{equation}
\label{hh}
\nu=\frac{1}{q_1\pm \frac{1}{ q_2\pm \frac{1}{q_3\pm\dots}}},\; q_1-odd,\; 
q_i-even\;integers,i=2,\dots,
\end{equation}
satisfactory reproduced experimentally observed fractional ratios, but many of them needed relatively high number of consecutive daughter quasiparticle or quasihole generations. This was regarded as a weakness of HH approach though it offers a systematic method for modeling of wave functions corresponding to particular fillings in the form of Laughlin-type condensate of quaiparticles or quasiholes created from the initial Laughlin state \cite{gr}. The relatively high step of consecutive daughter construction needed for many observed FQHE ratios in experiment meets with an inconvenient property related to the size of the following daughter quasiparticles or quasiholes, which quickly rises with the number of the generation (at the first generation this size is already of ca. ten magnetic length order, where the magnetic length $l_m=\sqrt{\frac{hc}{eB}}$) \cite{jain-n,jain2007}. The rapidly rising size of quasiparticles or quasiholes  would disturb formation of multiparticle correlated state with a high rank of the  daughter particle generation. 

The alternative phenomenological explanation of observed structure of FQHE in the  LLL was formulated by Jain and Wilczek in terms of so called composite fermions (CFs) \cite{jain} or superfermions \cite{wilczek}, employing idea of quantum statistics shift due to Aharonov-Bohm effect caused by auxiliary magnetic flux tubes pinned to exchanging particles. The development of the CF concept was done next by Read \cite{vor3} in terms of collective vortices reproducing Laughlin function for $\nu=\frac{1}{q}$ \cite{vor16}. The idea of Jain's CFs turned out to be very productive as allowed for explanation of the main line of experimentally observed FQHE filling factor hierarchy in the LLL \cite{jain,jain2007,hon}. It was achieved by remarkable observation that averaged field of CF flux tubes may reduce the external magnetic field and one can map FQHE onto IQHE of higher levels in weakened resultant field, successfully reproducing in this way majority of experimentally recorded ratios. Moreover, the CF idea allows for understanding of Hall metal states at some even denominator  filling fractions, when the total external magnetic field is canceled by averaged field of attached to particles flux-tubes. The idea of CFs was highly appealing as simple single-particle effective interpretation of strongly correlated multiparticle quantum fractional Hall state. Attachment of flux-tubes or vortices to electrons can be also easy represented in terms of appropriately rewritten or generalized Laughlin function \cite{jain2007,jain-n} and, on the other hand, by utilizing the scheme of 2D Chern-Simons field theory \cite{vor16,shankar1997,cs,lopez}. Exact diagonalizations confirmed related to CFs wave functions as a trial functions for FQH states from main line of hierarchy in the LLL and also used to model some selected states in higher LLs. 

Nevertheless, some problems  related to CF theory  can be specified as follows:
\begin{itemize}
\item {An artificial character of the construction of composite particles---there is  not specified any physical mechanism for creation of local auxiliary magnetic field flux tubes which can be fixed to moving particles. Moreover, in the framework of original CF formulation these pinned flux-tubes are of zero-dimension diameter---nowhere observed. There is not any microscopic mechanism which would produce from Coulomb interaction of electrons the quasiparticles dressed by such  magnetic field flux-tubes---the mass operator for interacting 2D electrons cannot produce such a quasiparticle (as a rule, quasiparticles are  defined by a pole of the retarded single-particle Green function \cite{abrikosov1975}). The interparticle separation quantization in 2D in magnetic field presence, as it was pointed out by Laughlin \cite{laughlin2,laughlin1} in terms of the  Coulomb interaction matrix element expressed in electron pair angular momentum quantum numbers, does not comply with the continuity of the mass operator requirement for the definition  of the Landau type quasiparticles   \cite{abrikosov1975}}.
\item{The failure of CF concept in exploration of fractional Hall states in the higher Landau levels---if CFs were true new particles (quasiparticles) arisen due to interaction they would manifest themselves also in the higher LLs, what is, however, in sharp contrast with the experimental data: in  the higher LLs FQHE is encountered extremely rarely in comparison to the  rich its manifestation in the LLL \cite{lls,ll8/3-1,5/2-1,ll8/3-2}. There is no rigorous idea based on CFs for modeling states in the higher LLs, thus it is carried out somewhat by tossing models and by further numerical verification of them trough comparison with the exact diagonalization; e.g., an idea of attachment of higher number (four) of flux tubes in higher LLs is not consistent with the quasiparticle concept of CFs (as leading to  different types of quasiparticles caused by the same dressing interaction).}
\item{An insufficiency in explanation of LLL ratios out of the main hierarchy (the main hierarchy is given by $\nu=\frac{n}{n(p-1)\pm1}$; the ratios out of the main hierarchy, but distinctly observed experimentally,  are e.g., $\nu=\frac{4}{11}, \frac{5}{13}, \frac{3}{10}, \frac{3}{8}, \frac{5}{7}, \frac{4}{5} $ and  other). Moreover, CF concept does not allow for an explanation of recent experiments in graphene revealing the triggering role of carrier mobility in formation of FQHE (the experiments with annealing of suspended graphene  \cite{fqhe1,bolotin2008,pfeiffer2003}), similarly as 
 of experiments in graphene \cite{fqhe1,fqhe2} demonstrating persistence of FQHE at lower magnetic field for more diluted (weaker interacting) electron gas.
An insufficiency occurs also in explanation of FQHE observation in bilayered 2DEG systems, especially of recent observation of FQHE in bilayered graphene \cite{2lg},}
\item{Not rigorously  established and in fact classical notion of a particle with flux-tube attached is linked rather with a single-particle picture but artificially conjectured to be equivalent to complex multiparticle correlated state. The irrelevance of CFs with magnetic field flux tubes appears in the case of fractional Chern insulators \cite{fci1,fci2,fci3,fci4} supporting analogy of FQHE without neither magnetic field nor LLs, but also exhibiting the  fractional filling hierarchy \cite{prodan,hasan2010,qi2011a,haldane5}.}
\item{If even one adopts the Read's concept of vortices attached to electrons instead of flux-tubes, one can observe that this multiparticle construction (vortices are collective objects \cite{vor3}) does not resolve the problem of CF origin---vortices are the fragments  of the Jastrow polynomial and the  vortices are in fact only decomposition of the Laughlin function with assumed in advance  (not derived) vorticity \cite{vor3,read1989}
without any justification for its value. 
The vorticity serves as equivalent phenomenological notion to the phase shift due to Aharonov-Bohm effect for flux tubes \cite{aharonow-bohm,wilczek}---both these approaches are effective illustrations of the Laughlin phase correlations but not the explanation. The same illustrative character has also implementation of CFs both with flux-tubes or vortices within Chern-Simons field theory, with suitably but artificially lifted singular gauge transformation to introduce flux-tubes or vortices CFs (but not serving as derivation of them) \cite{vor16,shankar1997,cs,lopez}; this gauge transformation (despite the name)  does not conserve the statistics and  changes the  type of particles, it is not a canonical transformation.}
\end{itemize}

The  wave functions of CFs base on the  reformulation of the Laughlin wave function  within the generalization of the scheme, $\prod_{i<j}(z_i-z_j)^q=\prod_{i<j}(z_i-z_j) \prod_{i<j}(z_i-z_j)^{q-1}$---the latter term in this expression represents attachment to fermions of $q-1$ flux-tubes (if to divide by its modulus) or vortex with $q-1$ vorticity, while the antisymmetric first factor can be modeled as variational trial function. It must be also noted, that
the quasiparticle concept of CFs conjectured by Jain   should not be confused with quasiparticles considered within HH approach. The HH concept for FQHE bases on fundamental observation of Laughlin \cite{laughlin2} that low energy excitations (called as quasiparticles or quasiholes) beyond or beneath the starting $\nu=\frac{1}{q}$ state originally described by the Laughlin function, are anyons with a fractional charge. Assuming then that these quasiparticles or quasiholes can themselves condensate into the next generation fractional state, one can get the hierarchy of occupation for LLL as given by Eq. (\ref{hh}). The daughter quasiparticles or quasiholes could again condensate and so on. These quasiparticles are not Jain's CFs which are conjectured as quite different quasiparticles  \cite{jain-n}.

Moreover, it is in order to note, that none of phenomenological explanations of FQHE (Jain's CFs, Read's CFs and HH model) gives arguments supporting the choice of the Laughlin function \cite{laughlin2,laughlin1} but rather oppositely---all employ known in advance Laughlin function to construct its illustration. The recently published results of exact diagonalization for 2D fractional Chern insulators \cite{fci1,fci2,fci3,fci4} strongly support an existence of more fundamental and universal origin of Laughlin correlations and fractional filling hierarchy exploiting the following prerequisite common for all related systems: sufficiently flat band, interaction and 2D geometry. The latter property seems to be underestimated in CF models in particular, and referred only to fractional anyonic statistics of low energy excitation beyond or beneath FQH state. In fact the specific 2D topology plays also central role in organization of FQH state itself. 
One can thus conclude that FQH features are the manifestation of a more fundamental multiparticle property of 2D interacting systems with some universal character not linked to magnetic field only, but also overwhelming the physics of Berry field flux quantization in fractional Chern insulators \cite{hasan2010}.

\section{Topological structure of CFs}
Some of listed above drawbacks of standard CF theory can be avoided if one look for topological reason of the phenomenological CF model. Such an outlook is possible if one employs  homotopy methods to description of many particle 2D charged system in the presence of quantizing perpendicular magnetic field strong enough to get fractional filling of the LLL. Similarly, one can consider 2D topological lattice supporting fractional Chern insulating state with magnetic field substituted by Berry field \cite{hasan2010}, flux of which is quantized in terms of Chern numbers.
Generally speaking the topology notions related to multiparticle system located at certain manifold, like on the plane in case of 2D systems, are purely classical referring to homotopy of multiparticle trajectories classified within the full braid group, i.e., the first homotopy group of the configuration space of $N$ identical indistinguishable particles \cite{artin1947,birman},
\begin{equation}
\pi_1(Q_N(M))=\pi_1(M^N\setminus\Delta)/S_N),
\end{equation}
where $\Delta $ is the diagonal subset of the configuration space $M^N$---the normal product of manifolds $M$ for each particle ($\Delta $ is subtracted in order to conserve the number of particles), $M=R^2$ for the plane, 
$S_N$ is the permutation group and the quotient structure is introduced for particle indistinguishability requirement. The space $Q_N(M)$ is not simply-connected thus its first homotopy group $\pi_1$ (the full braid group) is nontrivial: for $dim M \geq 3$ equals to finite permutation group, while for $dim M=2$ is infinite group defined by its generators $\sigma_i,\;i=1,2,\dots $ (exchanges of neighboring particles for selected their enumeration). The full braid group for the manifold $R^2$ may be formally defined by the following conditions imposed on its generators $\sigma_i$, $i=1,2,3,\dots N$ \cite{birman}:
\begin{equation}
\label{e10a}
\begin{array}{ll}
\sigma_i\sigma_{i+1}\sigma_i=\sigma_{i+1}\sigma_{i}\sigma_{i+1},\; & {\rm for}\; 1\leq i\leq N-2,\;\;\;\;\;\\
\end{array}
\end{equation}
\begin{equation}
\label{e10b}
\begin{array}{ll}
\sigma_i\sigma_j= \sigma_j\sigma_i.\; & {\rm for}\;1\leq i,j\leq N-1,\; |i-j|\geq 2.\\
\end{array}
\end{equation}

In order to connect the trajectory topology expressed in terms of braid groups with the quantum description of the considered multiparticle system the utilization of the path integral formalism is needed \cite{feynman1964,wilczek,wu,jac}. In Ref. \cite{lwitt} it is shown that in the case of not simply-connected configuration space, the path integral formula for the propagator $I_{a\rightarrow b}$ (it is a matrix element of the evolution operator in the position representation expressing probability of quantum transition between points $a$ and $b$ in the configuration space) has the form:
\begin{equation}
\label{f22}
I_{a\rightarrow b}=\sum\limits_{l\in\pi_1} e^{i\alpha_l}\int d\lambda_l e^{iS[\lambda_l(a,b)]/\hbar},
\end{equation}
$\pi_1$ is here the first homotopy group of the configuration space, i.e., the full braid group. $S[\lambda_l(a,b)]$  in the  exponent represents the action for  a trajectory $\lambda_l(a,b)$ linked $a$ and $b$ with attached a loop $l$ from $\pi_1$ (a trajectory $\lambda$ is open and an arbitrary loop $l$ can be added to it), $d\lambda_l$ represents here the measure in the homotopy sector $l$ of trajectories (in general, it cannot be defined  the same measure for the whole trajectory space due to discontinuity between various homotopy classes of $\pi_1$). The unitary factors, $e^{i\alpha_l}$, weigh the contributions of different homotopy classes of trajectories and in general they constitute a one-dimensional unitary representation (1DUR) of the full braid group $\pi_1$ \cite{lwitt}. Different choice of the representation defines a different type of quantum particles corresponding to the same classical multiparticle system.
In this way one can obtain bosons or fermions in the case of three-dimensional manifold where particles are located on. Indeed, for $dimM\geq 3$ there exist only two 1DURs of the permutation group (in 3D case the full braid group is the finite permutation group $S_N$, for the system of $N$ particles):
\begin{equation}
\sigma_l\rightarrow \left\{ \begin{array}{l}
e^{i0},\;\; bosons,\\
e^{i\pi},\;\;fermions,\\
\end{array}
\right.
\end{equation}
where $\sigma_l$ is the generator of the group $S_N$ (exchange of particles with numbers $l$ and $(l+1)$).
For two-dimensional manifolds the situation is more complicated because the full braid group is here an infinite group, not a permutation group. 
In the case of $M=R^2$, the infinite full braid group has also infinite number of distinct 1DURs, which for generators of the full braid group have the form:
\begin{equation}
\sigma_l\rightarrow e^{i\alpha},\;\alpha \in[0,2\pi),
\end{equation} 
note that the 1DUR element independence of index $l$ for $\alpha_l $ follows from the relation (\ref{e10a}) as 1DUR is Abelian. The corresponding particles are called anyons (including bosons for $\alpha =0$ and fermions for $\alpha=\pi$). 

Let us emphasize that the presented above analysis does not exhaust, however, all topological special features of $R^2$ manifold for multiparticle configuration space. The important property of the plane manifold manifests  itself in the presence of perpendicular to the plane magnetic field which is so strong that classical cyclotron trajectories are shorter in comparison to interparticle distances. If these distances are fixed in the uniform planar system of repulsing charged particles, then cyclotron trajectories cannot match neighboring particles. The braid trajectories must be built from cyclotron orbits at magnetic field presence, thus a problem with the definition of braid generators, $\sigma_i$, corresponding to exchanges of neighboring particles occurs. To restore particle exchanges the cyclotron orbits must somehow be enlarged. In standard CF theory this is achieved by reducing the field by auxiliary field flux tubes. In the case of vortices introduced by Read, the similar role plays screening of a local charge of the electron surrounded by the vortex (local dilution of electron density produces screening by positive jellium). Both these concepts model, however, the special topological and natural property of 2D manifold with charged particles exposed to perpendicular strong magnetic field. 

It has been demonstrated \cite{jac1,ws} that exclusively in 2D case substitution of singelooped trajectories by multilooped exchanges again allows for matching of neighboring particles. The braid generators $\sigma_i$ are substituted now by a new generators $(\sigma_i)^q, \; q - $odd integer (one can note that $(\sigma_i)^q=\sigma_i(\sigma_i)^{q-1}$ thus $(\sigma_i)^{q-1}$ represents $\frac{q-1}{2}$ additional loops). These new generators create the subgroup of the full braid group, which has been called 'cyclotron braid subgroup'. This braid subgroup is the appropriate $\pi_1$ group in the path integral (\ref{f22}). The 1DURs of the cyclotron braid subgroup reproduce the phase shift required by Laughlin correlations and it happens for LL filling factors $\nu=\frac{1}{q}$ \cite{jac1,ws}. Thus, it is shown that the Laughlin statistics can be obtained without invoking to artificial construction like CFs with auxiliary magnetic field flux-tubes attached to particles and reproducing the Laughlin statistics by Aharonov-Bohm phase shift. In view of the topological derivation of Laughlin correlations it is clear that CFs do not exist though are  the very convenient phenomenological effective single-particle model, however, with some limits of applicability which meets with the objections listed in the previous section.

It is worth noting the fact that when it comes to integration over trajectories, all possible trajectories contribute, regardless of dynamics details---but only if these trajectories are not excluded, what, however, happens at sufficiently strong magnetic field presence in two-dimensional interacting charged multiparticle systems. This a reason for emergence of the Laughlin correlations in FQHE.

The enhancement of cyclotron orbits in 2D due to adding of loops results from the property that on the plane any additional loop cannot change the surface of the planar system and all loops must share simultaneously the total flux of the external field and therefore the cyclotron radius is larger as accommodated to the fraction of this flux per each loop. The related commensurability condition for enlarged cyclotron orbits and interparticle separation allows for the  definition of the main FQHE hierarchy, 
\begin{equation}
\label{frac}
\nu = \frac{l}{l(q-1)\pm 1},
\end{equation} 
where $q$ is the number of loops ($q$ must be an odd integer because the related braids $(\sigma_i)^q$ must describe exchanges \cite{jac1,ws}), and $l 
=1,2,\dots$, whereas  $\pm$ corresponds to possible orientation of the last loop in the same direction or opposite to the orientation of the rest of loops (eight-figure-shape of trajectory in the case of opposite orientation) \cite{ws}. Assuming that the loops take away flux quanta, the last loop takes away the residual flux, in the case when it is impossible to divide the total flux onto integer number of quanta. If this residual flux fits to flux portion per particle in $l$-th LL (or at other filling corresponding to a correlated state), then the fraction (\ref{frac}) can be obtained. 

The multilooped braid explanation of FQHE \cite{jac2} agrees with recent experimental observations in suspended graphene \cite{fqhe1,fqhe2} demonstrating the triggering role of carrier mobility in formation of FQH state \cite{jac2013}. It has been shown that originally insulating state may be transformed into FQH state by annealing of the sample. Annealing reduces scattering of carriers and enhances mean free path of carriers without changing Coulomb interaction of electrons. Multilooped structure of orbits prefers longer mean free path of carriers. Thus, not only interaction and flat band are prerequisites for FQHE, but also sufficiently long mean free path of carriers is needed. Some additional support for significance of commensurability requirement follows from the observation of FQHE in graphene at low carrier concentration controlled by lateral voltage (thus with reduced interaction between diluted carries) and also at relatively lower magnetic field, when cyclotron orbits fit to larger interparticle separation. The most convincing arguments behind cyclotron braid group description of FQHE is the explanation of strikingly rare its manifestation in higher LLs in comparison to abundance of FQHE presence in the lowest LL. This will be illustrated in the following section.

\section{Explanation of hierarchy of quantum  Hall features  in the higher Landau levels} 

Very rich structure of FQHE in the lowest LL \cite{pan2003} remains in surprising opposition to its rather modest manifestation in higher LLs \cite{lls}. This phenomenon observed in many experiments reported e.g., in Refs \cite{lls,ll8/3-1,5/2-1,ll8/3-2,xia2004} had not found satisfactory explanation as of yet. It also does not agree with expectations based on the standard CF interpretation of FQHE \cite{jain,jain2007,hon}. 

The concept of cyclotron braid subgroups \cite{jac1,ws} allows for identification of LLL fillings at which the correlated multiparticle state FQHE can be arranged using the commensurability condition. Below we apply this to identification of FQHE-type and IQHE-type correlations in the system in several subbands of LLs. 

\subsection{Landau level structure}
Single-particle energy of LLs including spin  has the form ($g$--giromagnetic factor):
\begin{equation}
E_{n\uparrow(\downarrow)}=\hbar\omega_0\left(n+\frac{1}{2}\right) - (+)g\frac{1}{2}\hbar\omega_0,
\end{equation} 
where $\hbar\omega_0= \frac{\hbar eB}{mc}= 2\mu_B B $, $\mu_B=\frac{e\hbar}{2mc}$ is Bohr magneton.	
Thus the consecutive levels have the energy:
\begin{equation}
\begin{array}{l}
E_{0\uparrow}=\hbar\omega_0\left(\frac{1}{2}-\frac{g}{2}\right),\\
E_{0\downarrow}=\hbar\omega_0\left(\frac{1}{2}+\frac{g}{2}\right),\\
E_{1\uparrow}=\hbar\omega_0\left(\frac{3}{2}-\frac{g}{2}\right),\\
E_{1\downarrow}=\hbar\omega_0\left(\frac{3}{2}+\frac{g}{2}\right),\\
E_{2\uparrow}=\hbar\omega_0\left(\frac{5}{2}-\frac{g}{2}\right),\\
\dots\\
\end{array}
\end{equation}
The energy gaps separating following levels are:
\begin{equation}
g\hbar\omega_0,\;\;(1-g)\hbar\omega_0, \;\; g\hbar\omega_0,\;\; (1-g)\hbar\omega_0,\dots
\end{equation}
Degeneracy of each LLs is the same and depends on the magnetic field, 
\begin{equation}
N_0=\frac{BS}{hc/e},
\end{equation}
where $BS$ is the total flux of magnetic field $B$ trough the sample surface $S$, and $\Phi_0=\frac{hc}{e}$ is the quantum of magnetic field flux. 

Let us assume that the sample (2D) surface $S$ is constant as well as the constant number of electrons in the system $N$ is kept. Only the external magnetic field $B$ (perpendicular to the sample) can be changed resulting in the variation of LL degeneracy. 

\begin{figure}[ht]
\centering
\includegraphics{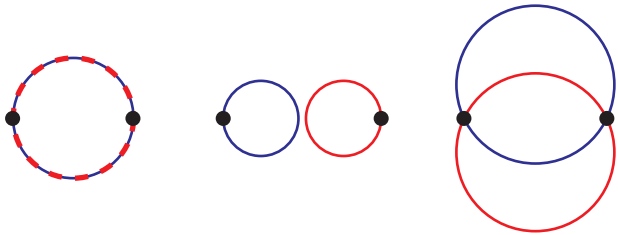}
\caption{Schematic demonstration that commensurabilty (left) of cyclotron orbit with interparticle separation satisfies topology requirements for braid interchanges in equidistantly uniformly distributed 2D particles; for smaller cyclotron radii particles cannot be matched (center), for larger ones the interparticle distance cannot be conserved (right) }
\label{fig:5}
\end{figure}

\begin{figure}[ht]
\centering
\includegraphics{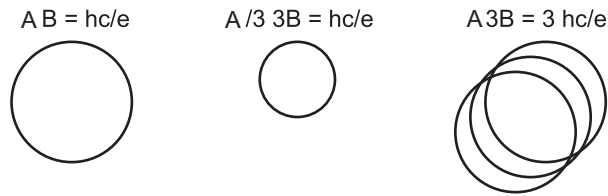}
\caption{Schematic illustration of cyclotron orbit enhancement in 2D due to multi-loop trajectory structure (third dimension added for visual clarity) }
\label{fig:6}
\end{figure}

To determine LL filling fractions corresponding to FQHE/IQHE collective states we will use the idea of commensurability of cyclotron orbits with interparticle 
separation in 2D system when Coulomb repulsion prohibits approaching one particle to another one. The commensurability condition displays ordering of the completely filled LLL, $\frac{hc}{eB_0}=\frac{S}{N}$. This condition means that the surface of the cyclotron orbit equals to the surface of the system per single particle. To imagine the link with braid group describing particles exchanges we notice that this condition allows existence of particle exchange trajectory along cyclotron classical orbits in the case of equidistantly and uniformly distributed particles possessing the same kinetic energy due to flatness of the band as in the case of LLs. The classical trajectory corresponding to the braid group generators and defined by cyclotron 2D orbits must fit to interparticle separations unless the statistics of particles could not be specified and a collective state would not form. The particularities of this approach are described in more details in Refs \cite{jac1,jac2,ws}. For shorthand let us refer here to the illustration in Fig. \ref{fig:5}, where the exchange trajectories for two particles are depicted in the topologically exclusive situation
of perfect fitting of cyclotron orbits to interparticle distance (left panel). For too short cyclotron trajectories the neighboring particles cannot be reached along these trajectories (central panel), for too large cyclotron orbits exchange trajectories cannot conserve the constant interparticle separation (right panel).   In the case of too short cyclotron orbits (as e.g., for LL filling factor $\nu=\frac{1}{3}$) in 2D system another possibility of exchanges occurs---along multilooped cyclotron orbits---cf. Fig. \ref{fig:6}. Because in 2D the surface of a planar orbit must be conserved thus a flux of the external field must be shared between all loops resulting therefore in reducing of the flux per each loop and effective enlarging of orbits. Multilooped structure of classical trajectories described by the cyclotron braid subgroup allows next for determination of quantum statistics, which satisfies Laughlin phase-correlation requirements \cite{jac1,jac2}. Thus, even though cyclotron orbit is a classical notion, the related its commensurability with particle distribution allows to resolve whether the collective multiparticle 2D state can be organized or not. This existential criterion will be applied in the following paragraphs to determine filling factor structure also of higher LLs. 

In Fig. \ref{fig:6} (left) the scheme of cyclotron orbit at magnetic field $B$ is shown as accommodated to the quantum of the  magnetic  field flux, i.e., $BA=\frac{hc}{e}$ (this is a definition of the cyclotron orbit $A$, and it is assumed that this $A$ fits to the interparticle separation, $A=\frac{ S}{N}$, $S$---the sample area, $N$---the number of particles). If only single-loop orbits are available, then at 3-times larger field, $3B$, the cyclotron orbit accommodated again to the flux quantum is too short in comparison to interparticle separation $\frac{S}{N}= A$ (which fits to the $B$ field orbits). This is illustrated in the central  panel of Fig. \ref{fig:6}. Nevertheless, if tree-loop orbits are considered, then in flat geometry of 2D space, the external flux $3BA$ must be shared between three loops  with the same surface $A$ (i.e., $BA$ for each loop).  Thus, each loop accommodated to flux quantum $\frac{hc}{e}$ has  the orbit with the surface $A$ and gives the contribution $BA$ to the  flux, resulting in the total flux $3BA$ per particle, as needed---this is schematically illustrated in Fig. \ref{fig:6} (right). The size of $A$ in the right panel is equal to $A$ in the left panel, which means that the  three-loop orbits fit to the interparticle separation defined by $A$.  

\subsubsection{Completely filled subband $0\uparrow$ of the LLL} 

If $B=B_0$, when $N_0=N$, i.e., $N=\frac{B_0 S}{hc/e}$, then the filling factor $\nu =\frac{N}{N_0}=1$ and we deal with the completely filled lowest subband of the zeroth LL exhibiting fully developed IQHE for spin-polarized electrons.

\subsubsection{Fractional ($1/q$, $q$-odd) fillings of $0\uparrow$ subband of the LLL}
	
For higher magnitudes of the field, $ B>B_0$ we have $N_0>N$, i.e., $\nu=\frac{N}{N_0}<1$. In this case the cyclotron single-loop orbits are too short to reach neighboring particles and in order to organize a collective state the multilooped cyclotron orbits are required. In 2D additional loops do not change the surface area and the external field flux must be shared by all loops resulting in enhancement of their size. Thus e.g., for $B=3B_0$, the singlelooped cyclotron trajectories do not reach neighboring particles, while the threelooped trajectories fit accurately to interparticle distances. For $q$-looped orbits the commensurability condition, 
\begin{equation}
q\frac{hc}{eB}=\frac{S}{N},
\end{equation}
gives 
\begin{equation}
\label{fr0}
\nu=\frac{N}{N_0}=\frac{SeB/qhc}{SeB/hc}=\frac{1}{q}.
\end{equation}
Because $q$ must be odd to prevent that corresponding exchange trajectories are braids \cite{jac2} we get in this manner the main ratios for FQHE, $\nu=\frac{1}{q}$. All electrons are in the same subband, $0\uparrow$ of LLL, thus the effect is strong, which is visible in experiments. 

One-dimensional unitary representation of cyclotron braid subgroup gives the phase shift when particles exchange. This particle exchange is understand as an exchange of arguments of the multiparticle wave function. In 2D, the  exchanges of arguments of the wave function are displayed by braid group elements distinct than the simple permutation as in 3D. The one-dimensional unitary representation of the particular braid gives then the phase shift corresponding to the chosen exchange \cite{imbo1988,sud} in perfect coincidence with the Laughlin correlations.

\subsubsection{Even or odd denominator filling ratios in the subband $0\uparrow$}

In the subband $0\uparrow$ cyclotron orbits are of size, $\frac{hc}{eB}=\frac{S}{N_0}$, and always are smaller than $\frac{S}{N}$, as $N<N_0$ (i.e., $\nu<1$). As it was described above, the additional loops are thus needed to match neighboring particles along cyclotron braids, i.e., $
\frac{q hc}{eB}=\frac{S}{N}, \;q-odd,\;\rightarrow \nu=\frac{1}{q}$.
Some other fractions can be determined by commensurability condition for the last loop of multilooped orbit, when the former loops takes the flux quantum each. In this way the filling fractions which are  listed  in the Tab. \ref{tab2}, according to the hierarchy,
\begin{equation}
\label{fr1}
\nu=\left\{  
\begin{array}{l}
 \frac{l}{l(q-1)\pm 1},\; for\; electrons,\\
1-\frac{l}{l(q-1)\pm 1},\; for\; holes,\\
\end{array}
\right.
\end{equation}
can be achieved, where $l$ may be taken equal to filling ratio of other correlated state including completely filled subbands, cf also \cite{uwaga}.

\begin{table}[h!]
\centering
\begin{tabular}{|p{3.5cm}|p{1cm}|p{1cm}|p{1cm}|p{1cm}|p{1.2cm}|}
\hline
$l$ and $\pm$ in Eq. (\ref{fr1})&$q=3$&$q=5$&$q=7$&$q=9$&$q=11$\\
\hline
1 $(+)$&1/3&1/5&1/7&1/9&1/11\\
\hline
1 $(-)$&1&1/3&1/5&1/7&1/9\\
\hline
2 $(+)$&2/5&2/9&2/13&2/17&2/21\\
\hline
2 $(-)$&2/3&2/7&2/11&2/15&2/19\\
\hline
3 $(+)$&3/7&3/13&3/19&3/5&3/31\\
\hline
3 $(-)$&3/5&3/11&3/17&3/23&3/29\\
\hline
4 $(+)$& 4/9&4/17&4/25&4/33&4/41\\
\hline
4 $(-)$&4/7&4/15&4/23&4/31&4/39\\
\hline
5 $(+)$&5/11&5/21&5/31&5/41&5/51\\
\hline
5 $(-)$&5/9&5/19&5/29&5/39&5/49\\
\hline
6 $(+)$&6/13&6/25&6/37&6/49&6/61\\
\hline
6 $(-)$&6/11&6/23&6/35&6/47&6/59\\
\hline
7 $(+)$&7/15&7/29&7/43&7/57&7/71\\
\hline
7 $(-)$&7/13&7/27&7/41&7/55&7/69\\
\hline
8 $(+)$&8/17&8/33&8/49&8/65&8/81\\
\hline
8 $(-)$&8/15&8/31&8/47&8/63&8/79\\
\hline
9 $(+)$&9/19&9/37&9/55&9/73&9/91\\
\hline
9 $(-)$&9/17&9/35&9/53&9/71&9/89\\
\hline
10 $(+)$&10/21&10/41&10/61&10/81&10/101\\
\hline
10 $(-)$&10/19&10/39&10/59&10/79&10/99\\
\hline
\hline
4/3 $(+)$&4/11&4/19&4/27&4/35&4/43\\
\hline
4/3 $(-)$&4/5&4/13&4/21&4/29&4/37\\
\hline
5/3 $(+)$&5/13&5/23&5/33&5/43&5/53\\
\hline
5/3 $(-)$&5/7&5/17&5/27&5/37&5/47\\
\hline
6/5 $(+)$&6/17&6/29&6/41&6/53&6/65\\
\hline
6/5 $(-)$&6/7&6/19&6/31&6/43&6/55\\
\hline
9/5 $(+)$&9/23&9/41&9/59&9/77&9/95\\
\hline
9/5 $(-)$&9/13&9/31&9/49&9/67&9/85\\
\hline
7/3 $(+)$&7/17&7/31&7/45&7/59&7/73\\
\hline
7/3 $(-)$&7/11&7/25&7/39&7/53&7/67\\
\hline
8/3 $(+)$&8/19&8/35&8/51&8/67&8/83\\
\hline
8/3 $(-)$&8/13&8/29&8/45&8/61&8/77\\
\hline
3/2 $(+)$&3/8&3/14&3/20&3/26&3/32\\
\hline
3/2 $(-)$&3/4&3/10&3/16&3/22&3/28\\
\hline
\end{tabular}
\caption{The LLL filling factors for FQHE determined by commensurability condition (acc. to the formula (\ref{fr1}), for electrons only)}
\label{tab2}
\end{table}

If for the last loop falls zero flux, i.e., the former $q-1$ loops take away the total flux of $B$ field, then the last loop describes free movement as in the Fermi liquid without the field. This happens for $\nu=\frac{1}{q-1}=\frac{1}{2},\frac{1}{4} \dots$ and symmetrically for holes,
$\nu=1-\frac{1}{q-1}=\frac{3}{4},\frac{5}{6},\dots$. The corresponding Hall states are referred as to Hall metal states \cite{kr,kr1,pasquier1998,papic2009,ssw1,vor3,halperin,vor7,vor11,vor9}. Note that these filling ratios correspond to the formula (\ref{fr1}) taken in the limit $l\rightarrow \infty $ (then for the last loop falls zero flux as per particle in $l$th LL with $l\rightarrow \infty$).

\subsubsection{Fillings of the subband $0\downarrow$ of the LLL}

Now let us consider $B<B_0$, that $2N_0>N>N_0$. Then the subband $0\uparrow$ is completely filled with $N_0$ electrons, while the rest of electrons, $N-N_0<N_0$ must be located in $0\downarrow$ subband. For electrons $0\uparrow$ their cyclotron orbits equal to $\frac{hc}{eB}=\frac{S}{N_0}$ and these electrons form a collective state of IQHE. The electrons $0\downarrow$ (they are distinct from the previous ones) fill the subband $0\downarrow$ and their cyclotron orbits $\frac{hc}{eB}$ are certainly shorter in comparison to particle separation $\frac{S}{N-N_0}$. The multilooped trajectories are thus needed corresponding to FQHE for this electrons, similarly as in the subband $0\uparrow$ .

For $q$-looped ($q$-odd to assure particle exchanges),
\begin{equation}
q\frac{hc}{eB}=\frac{S}{N-N_0},
\end{equation}
thus $\nu=\frac{N}{N_0}=1+\frac{1}{q}=\frac{4}{3}, \frac{6}{5},\frac{8}{7},\dots$
and dually for holes in this subband, $\nu =1+\frac{q-1}{q}=\frac{5}{3},\frac{9}{3}, \frac{13}{7}, \frac{17}{9},\dots$ . In both these cases in the subbband $0\downarrow$ we have less than half of total number of electrons---thus corresponding to them FQHE might be not so pronounced in comparison to $0\uparrow$ subband, and accompanied by IQHE of completely filled lower subband.

Here one can
again consider the situation when $q-1$ loops take away the flux quantum each, while for the last loop a fraction of flux quantum falls, in such a way that this fraction corresponds to the fraction of flux quantum per particle in higher LLs (or at other correlated state filling). This leads to possible fractions, due to commensurability condition fulfilled, 
\begin{equation}
\label{fr2}
\nu=\left\{  
\begin{array}{l}
 1+\frac{l}{l(q-1)\pm 1},\; for\; electrons,\\
2-\frac{l}{l(q-1)\pm 1},\; for\; holes,\\
\end{array}
\right.
\end{equation}
where $l$ may be taken equal to the filling fraction  for other correlated state including completely filled subbands.

In the case when $q-1$ loops take away the total flux of $B$ field, the last loop describes field-free movement as in Fermi liquid without the field presence.
This happens for $\nu=\frac{3}{2},\frac{5}{4}, \frac{7}{4}\dots$ (Hall metal states in $0\downarrow$ subband, at fractional fillings corresponding to  the limit $l\rightarrow \infty$ of Eq. (\ref{fr2})).

The structure of Hall features in the subband $0\downarrow$ repeats that one from the subbband $0\uparrow$, with only shift by 1 in  $\nu$. The same symmetry will be hold for the spin-doublets of subbands in all higher LLs. This is illustrated in Fig. \ref{fig:1}.

\subsubsection{Completely filled both subbands $0\uparrow$ and $0\downarrow$}
For lower field $B$, that $N=2N_0$, the cyclotron orbits in both subbands $0\uparrow$ and $0\downarrow$ ideally fit to interparticle separations of two sorts of particles $0\uparrow$ and $0\downarrow$, i.e., $\frac{hc}{eB}=\frac{S}{N_0}=\frac{S}{N/2}$, and in both subbands of the LLL we deal with IQHE.

\subsubsection{Fillings of $n=1$ LL, the subband $1\uparrow$}

If now the magnetic field is still lowering, that $3N_0>N>2N_0$, then the subbands of LLL, $0\uparrow$ and $0\downarrow$, are completely filled with IQHE for both types of electrons, while the rest of electrons, i.e., $N-2N_0$ is located in the subband $1\uparrow$.

The cyclotron orbit from this subband equals to $\frac{3hc}{eB}$ (note that the LL energy without spin, i.e., the kinetic energy of flat band decides on the size of cyclotron radius at field $B$) and its radius may ideally fit to interparticle separation in this subband $\frac{S}{N-2N_0}$, or can  be lower than separation, or even higher than this separation. The latter  is the new situation in comparison to both subbands of $n=0$ LLL. 
Thus there are now the following possibilities:
\begin{enumerate}
\item{The cyclotron orbit from this subband equals to $\frac{3hc}{eB}$ (let us remind  that the LL energy without spin, i.e., the kinetic energy of flat band defines the size of the cyclotron radius at field $B$) and  it dimension may ideally fit to interparticle separation in this subband, then 
\begin{equation} 
\frac{3hc}{eB}=\frac{S}{N-2N_0},
\end{equation}
thus $\nu=\frac{N}{N_0}=\frac{7}{3}$ because $\frac{3hc}{eB}=\frac{3S}{N_0}=\frac{S}{N-2N_0}$, and in all subbands we deal with IQHE as all cyclotron orbits are singlelooped, though the subband $1\uparrow$ is not completely filled  (the correlated states resembling IQHE but for not completely filled subbands we will denote as IQHEc).}
\item{The another situation is when the cyclotron orbit $\frac{3hc}{eB}$ in this subband is too short to match particles in this subband, $ \frac{3hc}{eB}<\frac{S}{N-2N_0}$. Then the multilooped ($q$-looped, $q$-odd) trajectories are needed resulting then in FQHE for this electrons,
\begin{equation}
q\frac{3hc}{eB}=\frac{S}{N-2N_0},
\end{equation}
then $\nu = \frac{6q+1}{3q}=\frac{19}{9}, \frac{31}{15}, \frac{43}{21},\dots $, and the corresponding FQHE is weak, as concerns only lower than 1/3 of electrons. 
}
\item{	The next possibility for commensurability of cyclotron orbits and interparticle separation of electrons from the subband $1\uparrow$ is in the case when the cyclotron orbit is twice in dimension of interparticle separation:
\begin{equation}
\frac{3hc}{eB}=\frac{2S}{N-2N_0},
\end{equation}
then $\nu=\frac{8}{3}$, because in this case $\frac{3S}{N_0}=\frac{2S}{N-2N_0}$. In this case we deal with IQHE (with singlelooped orbits) in all subbands including the last one incompletely filled (IQHEc), however.} 
\item{For lower field $B$, when the cyclotron orbit in the subband $1\uparrow$,
$\frac{3hc}{eB}$, fits to three interparticle separations, $\frac{S}{N-2N_0}$, i.e.,
\begin{equation}
\frac{3hc}{eB}=\frac{3S}{N-2N_0}, 
\end{equation}
then $N=3N_0$, and $\nu = \frac{N}{N_0}=3$, which corresponds to completely filled three first subbands, $0\uparrow$, $0\downarrow$ and $1\uparrow$, with IQHE correlation in each subband.}
\item{The special filling corresponds to the commensurabilty condition:
\begin{equation}
\label{pair}
\frac{3hc}{eB}=\frac{3S}{N_0}=\frac{1.5S}{N-2N_0}\rightarrow \nu=\frac{5}{2}. 
\end{equation}
If one assumes that electrons from the subband $1\uparrow$ would create pairs (BCS-like state described by Pfaffian), then total number of particle pairs in this subband equals to $\frac{N-N_0}{2}$ what gives from condition (\ref{pair}) the ideal fitting to cyclotron orbits as for IQHEc, but for pairs instead of electrons (note that the cyclotron orbit radius scales as $\sim \frac{e}{m}=\frac{2e}{2m}$ and does not change for electron pairs). 
Apparently the state corresponding to $\nu=\frac{5}{2}$ does not repeat the structure of states for $\nu=\frac{3}{2},\frac{1}{2}$ of Hall metal type, which agrees with experimental observations and numerical modeling \cite{5/2-1}.}
\end{enumerate} 

With regard to FQHE in the subband $1\uparrow$ one can consider that $q-1$ loops had taken the flux quantum each, while the last loop, the rest of the total flux of the external field. If this fraction fits to flux per particle in higher LLs or at other correlate state filling, the new ratios are available. The same with the situation when this rest is zero, which creates opportunity to next even denominator fractions in the subband $1\uparrow$---these onces in analogy to fractions $\frac{1}{2}, \frac{1}{4}$ with Hall metal manifestation.

\subsubsection{Fillings of the subband $1\downarrow$ of the $n=1$ LL}
When we still lower the magnetic field $B$ we attain the region $4N_0>N>3N_0$, corresponding to filling of $1\downarrow$ subband. In this case three antecedent subbands $0\uparrow,\;0\downarrow,\;1\uparrow$ are completely filled. In subbands $0\uparrow,\;0\downarrow$ cyclotron orbits are $\frac{hc}{eB}$ and they are ideally commensurate with interparticle separation in these subbands equal to $\frac{S}{N_0}$. Thus the correlation in this subbands corresponds to IQHE.

In the subband $1\uparrow$ the cyclotron orbit has a surface $\frac{3hc}{eB}$ three times larger than $\frac{S}{N_0}$. Due to this commensurability condition the exchanges of every third electron allow for IQHE on such subset of electrons. 
In subband $1\downarrow$ we have thus $N-3N_0<N_0$ electrons. The cyclotron orbit $\frac{3hc}{eB}$ may be here too small, equal or too large in comparison to $\frac{S}{N-3N_0}$.
\begin{enumerate}
\item{If this cyclotron orbit is too small for interchanges, then multilooped trajectories are required resulting in FQHE of related electrons. This happens at $\nu=\frac{N}{N_0}=\frac{9q+1}{3q}=\frac{28}{9},\frac{46}{15}, \frac{64}{21},\dots$ (because $\frac{3qS}{eB}=\frac{S}{N-3N_0}$, $q$-odd). In this case the correlation in $1\downarrow$ subband has FQHE character.
The symmetric ratios for holes in the subband $1\downarrow$ must be also taken into account.}
\item{If $\frac{3hc}{eB}=\frac {3S}{N_0}=\frac{S}{N-3N_0}$, then $\nu =\frac{10}{3}$ and the corresponding correlation in $1\downarrow$ has IQHEc character.}
\item{For $\frac{3hc}{eB}=\frac{3S}{N_0}=\frac{2S}{N-3N_0}$, i.e., for $\nu=\frac{N}{N_0}=\frac{11}{3}$, the commensurability holds for every second electron in the subband resulting in IQHEc correlation for electron subset in this subband. }
\item{Finally, for $\frac{3hc}{eB}=\frac{3S}{N_0}=\frac{3S}{N-3N_0}$ we have $\nu=4$ and IQHE corresponding to commensurability of cyclotron orbit with every third electron in the subband $1\downarrow$.}
\item{In the subband $1\downarrow$ the commensuarability condition,
\begin{equation}
\frac{5hc}{eB}=\frac{2.5S}{N-3N_0}\rightarrow \nu=\frac{7}{2},
\end{equation}
similarly as for filling fraction $\nu=\frac{5}{2}$, gives opportunity for paring of electrons and ideal fitting of cyclotron braids for pairs as in IQHEc.}
\end{enumerate}

In this subband also for FQHE opens possibility to get ratios by accommodation of the residual fraction of flux quantum piercing the las loop of $q$-looped cyclotron orbits with the fraction of flux per particle in higher LLs (or at other selected filling). In particular one can consider also this rest flux to be equaled zero in analogy to $\frac{1}{2}$ filling (Hall metal).

\subsubsection{Fillings of $n=2$ LL, the subband $2\uparrow$}
	For sufficiently low magnetic field, that $5N_0>N>4N_0$, i.e., $\nu \in (4,5]$ we deal with four first LL subbands completely filled. In subbands $0\uparrow$ and $0\downarrow$ the cyclotron orbits $\frac{hc}{eB}$ fit to interparticle separation, i.e., $\frac{hc}{eB}=\frac{S}{N_0}$. In subbands $1\uparrow$ and $1\downarrow$ the commensurability condition has the form $\frac{3hc}{eB}=\frac{3S}{N_0}$. 
The remaining electrons fill now the subband $2\uparrow$.

In the subband $2\uparrow$ are $N-4N_0<N_0$ electrons and cyclotron orbits have the size $\frac{5hc}{eB}$.

\begin{enumerate}
\item{If cyclotron orbits in $2\uparrow$ fit to interparticle separation, 
\begin{equation}
\frac{5hc}{eB}=\frac{S}{N-4N_0},
\end{equation}
then $\nu=\frac{21}{5}$ corresponding to IQHE in all filled subbands and IQHEc in fractionally filled $2\uparrow$ subband.}
\item{For cyclotron orbit shorter than separation of particles in the subband $2\uparrow$ the multilooped cyclotron orbits are required resulting in FQHE in this subband. This happens for filling ratios: 
\begin{equation}
\begin{array}{l}
\frac{q5hc}{eB}=\frac{S}{N-4N_0}
=\frac{q5S}{N_0} \rightarrow 
\nu=\frac{20q+1}{5q}=\frac{61}{15},\frac{101}{25},\frac{141}{35},\dots
\end{array}
\end{equation}
corresponding to FQHE in the subband $2\uparrow$.
The symmetrical filling ratios (versus the center of the subband) can be additionally associated with holes in the subband. }
\item{For commensurability condition $\frac{5hc}{eB}=\frac{2S}{N-4N_0}$, $\nu=\frac{22}{5}$, corresponding to IQHEc in the $2\uparrow$ subband.}
\item{For commensurability condition $\frac{5hc}{eB}=\frac{3S}{N-4N_0}$, $\nu=\frac{23}{5}$, corresponding to IQHEc in the $2\uparrow$ subband.}
\item{For commensurability condition $\frac{5hc}{eB}=\frac{4S}{N-4N_0}$, $\nu=\frac{24}{5}$, corresponding to IQHEc in the $2\uparrow$ subband.}
\item{Finally, for commensurability condition $\frac{5hc}{eB}=\frac{5S}{N-4N_0}$, $\nu=5$ and we deal with IQHE of completely filled first five LL subbands.}
\item{The special commensurability condition,
\begin{equation}
\frac{5hc}{eB}=\frac{2.5S}{N-4N_0}\rightarrow \nu=\frac{9}{2},
\end{equation}
give opportunity for pairing of electrons in the subband $2\uparrow$.}
\end{enumerate}

\subsubsection{Fillings of $n=2$ LL, the subband $2\downarrow$}
Again lowering the field one arrives at the region 
$6N_0>N>5N_0$, i.e., $\nu \in (5,6]$ corresponding to five first LL subbands completely filled and partially filled $2\downarrow$ subband. In subbands $0\uparrow$ and $0\downarrow$ the cyclotron orbits $\frac{hc}{eB}$ fit to interparticle separation, i.e., $\frac{hc}{eB}=\frac{S}{N_0}$. In subbands $1\uparrow$ and $1\downarrow$ the commensurability condition has the form $\frac{3hc}{eB}=\frac{3S}{N_0}$ as usual for completely filled $n=1$ LL. The subband $2\uparrow $ is also completely filled with the commensurate condition
$\frac{5hc}{eB}=\frac{5S}{N_0}$. In all these fully filled subbands a correlated state is of the IQH type.
The remaining electrons fill now the subband $2\downarrow$.
In the subband $2\downarrow$ are $N-5N_0<N_0$ electrons and cyclotron orbits have the size $\frac{5hc}{eB}$.
There are following possibilities here:
\begin{enumerate}
\item{If cyclotron orbits in $2\downarrow$ fit to interparticle separation, 
\begin{equation}
\frac{5hc}{eB}=\frac{S}{N-5N_0},
\end{equation}
then $\nu=\frac{26}{5}$ corresponding to IQHE in all filled subbands and IQHEc in fractionally filled $2\downarrow$ subband.}
\item{For cyclotron orbit shorter than separation of particles in the subband $2\downarrow$ the multilooped cyclotron orbits are required resulting in FQHE in this subband. This happens for filling rations: 
\begin{equation}
\begin{array}{l}
\frac{q5hc}{eB}=\frac{S}{N-5N_0}
=\frac{q5S}{N_0} \rightarrow 
\nu=\frac{25q+1}{5q}=\frac{76}{15},\frac{126}{25},\frac{186}{35},\dots
\end{array}
\end{equation}
corresponding to FQHE in the subband $2\downarrow$.}
\item{For commensurability condition $\frac{5hc}{eB}=\frac{2S}{N-5N_0}$, $\nu=\frac{27}{5}$, corresponding to IQHEc in the $2\downarrow$ subband.}
\item{For commensurability condition $\frac{5hc}{eB}=\frac{3S}{N-5N_0}$, $\nu=\frac{28}{5}$, corresponding to IQHEc in the $2\downarrow$ subband.}
\item{For commensurability condition $\frac{5hc}{eB}=\frac{4S}{N-5N_0}$, $\nu=\frac{29}{5}$, corresponding to IQHEc in the $2\downarrow$ subband.}
\item{Finally, for commensurability condition $\frac{5hc}{eB}=\frac{5S}{N-5N_0}$, $\nu=6$ and we deal with IQHE of completely filled first six LL subbands.}
\item{The special commensurability condition,
\begin{equation}
\frac{5hc}{eB}=\frac{2.5S}{N-5N_0}\rightarrow \nu=\frac{11}{2},
\end{equation}
due to  pairing of electrons at $\nu=\frac{11}{2}$ in the subband $2\downarrow$.}
\end{enumerate}

\subsubsection{Fillings of $n=3$ LL, the subband $3\uparrow$}
In order to consider $n=3$ LL the field is lowering once more to attain the region 
$7N_0>N>6N_0$, i.e., $\nu \in (6,7]$ corresponding to six first LL subbands completely filled and partially filled $3\uparrow$ subband. In subbands $0\uparrow$ and $0\downarrow$ the cyclotron orbits $\frac{hc}{eB}$ fit to interparticle separation, i.e., $\frac{hc}{eB}=\frac{S}{N_0}$. In subbands $1\uparrow$ and $1\downarrow$ the commensurability condition has the form $\frac{3hc}{eB}=\frac{3S}{N_0}$ corresponding to fully filled $n=1$ LL. The subbands $2\uparrow $ and $2\downarrow$ are also completely filled with the commensurate condition
$\frac{5hc}{eB}=\frac{5S}{N_0}$. In all these filled subbands a correlated state is of the IQH type.
The remaining electrons fill now the subband $3\uparrow$.
In the subband $3\uparrow$ are $N-6N_0<N_0$ electrons and cyclotron orbits have the size $\frac{7hc}{eB}$.
There are following possibilities here:
\begin{enumerate}
\item{If cyclotron orbits in $3\uparrow$ fit to interparticle separation, 
\begin{equation}
\frac{7hc}{eB}=\frac{S}{N-6N_0},
\end{equation}
then $\nu=\frac{43}{7}$ corresponding to IQHE in all filled subbands and IQHEc in fractionally filled $3\uparrow$ subband.}
\item{For cyclotron orbit shorter than separation of particles in the subband $3\uparrow$ the multilooped cyclotron orbits are required resulting in FQHE in this subband. This happens for filling ratios: 
\begin{equation}
\begin{array}{l}
\frac{q7hc}{eB}=\frac{S}{N-6N_0}
=\frac{q7S}{N_0} \rightarrow 
\nu=\frac{42q+1}{7q}=\frac{127}{21},\frac{211}{35},\frac{295}{49},\dots
\end{array}
\end{equation}
corresponding to FQHE in the subband $3\uparrow$.}
\item{For commensurability condition $\frac{7hc}{eB}=\frac{2S}{N-6N_0}$, $\nu=\frac{44}{7}$, corresponding to IQHEc in the $3\uparrow$ subband.}
\item{For commensurability condition $\frac{7hc}{eB}=\frac{3S}{N-6N_0}$, $\nu=\frac{45}{7}$, corresponding to IQHEc in the $3\uparrow$ subband.}
\item{For commensurability condition $\frac{7hc}{eB}=\frac{4S}{N-6N_0}$, $\nu=\frac{46}{7}$, corresponding to IQHEc in the $3\uparrow$ subband.}
\item{For commensurability condition $\frac{7hc}{eB}=\frac{5S}{N-6N_0}$, $\nu=\frac{47}{7}$, corresponding to IQHEc in the $3\uparrow$ subband.}
\item{For commensurability condition $\frac{7hc}{eB}=\frac{6S}{N-6N_0}$, $\nu=\frac{48}{7}$, corresponding to IQHEc in the $3\uparrow$ subband.}
\item{Finally, for commensurability condition $\frac{7hc}{eB}=\frac{7S}{N-6N_0}$, $\nu=7$ and we deal with IQHE of completely filled first seven LL subbands.}
\item{The special commensurability condition,
\begin{equation}
\frac{7hc}{eB}=\frac{3.5S}{N-6N_0}\rightarrow \nu=\frac{13}{2},
\end{equation}
 corresponds to pairing of electrons in the subband $3\uparrow$ at the above filling rate.
Similarly for the next subband $3\downarrow$, 
the commensurability condition,
\begin{equation}
\frac{7hc}{eB}=\frac{3.5S}{N-7N_0}\rightarrow \nu=\frac{15}{2},
\end{equation}
results in pairing of electrons in the subband $3\downarrow$ at $\nu=\frac{15}{3}$.}
\end{enumerate}

\begin{figure}[h]
\centering
\scalebox{1.8}{\includegraphics{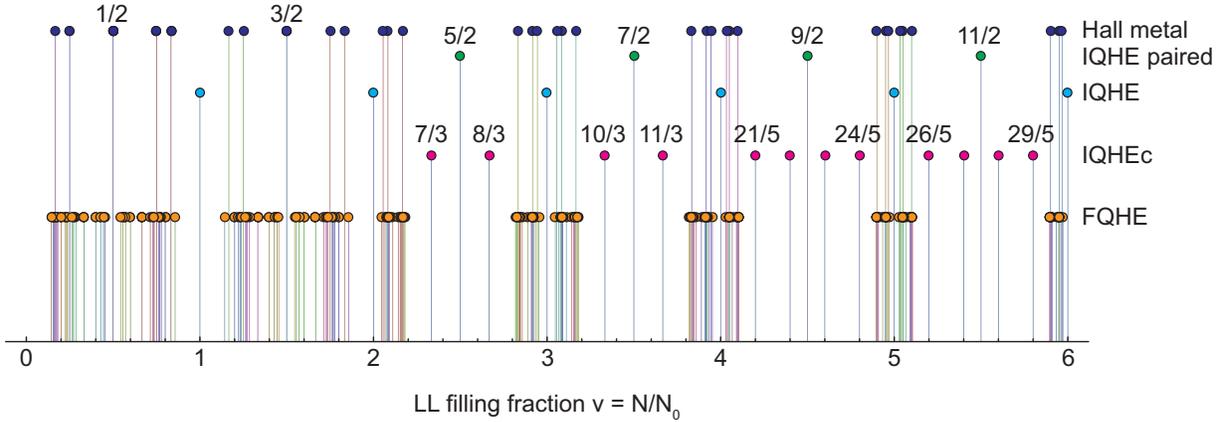}}
\caption{Graphical presentation of filling factors up to sixth LL subband selected by commensurability condition: each spike represents filling ratio for possible correlated state, the lowest spikes correspond to series of FQHE hierarchy given by (\ref{fr0}), (\ref{fr1}), (\ref{fr2}) and (\ref{fr3}), the  taller correspond  to IQHEc inside bands with $ n\geq 1$, the tallest ones correspond to paired states acc. (\ref{frac3}) and Hall metal states acc. (\ref{fr4}) or (\ref{fr1}) and (\ref{fr2}) taken in the limit $ l\rightarrow \infty$;    the evolution of FQHE fillings is visible with increasing  number of LL subband; IQHEc indicates states selected by commensurability condition corresponding to single-loop cyclotron orbits as in completely filled subband but not protected by large inter-level energy gap; IQHE paired indicates states with pairing preferable by commensurability}
\label{fig:1}
\end{figure}

Described above structure of filling ratios for LLs is summarized in Tab. \ref{tab1} and schematically illustrated in Fig. \ref{fig:1}.

\section{Summarizing of filling ratios corresponding to cyclotron orbit commensurability with interparticle separation}

In the similar manner as presented above, one can analyze fillings of spin  subbands of next LLs and determine the special filling hierachy corresponding to organization of collective states of FQHE or IQHEc-type in consecutive LLs. The role of interaction is crucial here---due to the Coulomb repulsion the interparticle separation can be kept fixed in the system. Besides the interaction the two-dimensionality of the system is the important prerequisite. The multilooped structure of braids resulting in FQHE is possible in 2D exclusively. The hierarchy of states resulting from the commensurability of cyclotron orbits with interparticle separation reflects possibility of collectivization of multiparticle system by particle exchanges which requires topologically admitted braids, thus necessity for fitting of cyclotron orbits to particle separation. If this fitting is missed, the braid structure cannot be defined for the system. 

Nevertheless, due to  definition of braid generators by multilooped trajectories  the additional possibility occurs associated with FQHE.  If in the case of multilooped trajectories each loop takes away single flux quantum, one arrives at $\nu=\frac{1}{p}$ filling ratios for FQHE. However, when the last loop of multilooped trajectory takes away the residual fraction of flux quantum the same as formally calculated for $l$ filling fraction of other correlated state (in particular, for completely filled subbands), then the hierarchy (\ref{frac})
can be established in  correspondence with the CF idea formulated for the LLL.  The sign minus in the above formula is linked to the possibility of inverted direction of the last loop (eight-figure shape of the trajectory), in analogy to the possibility of inverted orientation of resultant field screened by flux tubes within the CF concept. Some part of the resulting fractions is listed in Tab. \ref{tab2}.

For the higher LLs the situation is more complicated as it has been illustrated in the above paragraphs up to $n=3$. The formula (\ref{fr1}) can be generalized to the higher LL case. The related derivation is placed  in the next paragraph. The generalized formula (\ref{fr3})  corresponds to the FQHE states in the higher LLs. Similarly, its limit $l\rightarrow \infty$, i.e., the formula (\ref{fr4}) defines fractions for Hall metal states.  Besides these features for LL with $\geq 1$ the new opportunities occur, not possible for $n=0$. They correspond to single-loop commensurability similar as for IQHE and are referred to correlated states called as IQHEc with ordering similar to IQHE but for not completely filled subband.    The all resulting filling ratios are included in the summarizing Table \ref{tab1}. In this table there are summarized filling ratios for FQHE and IQHE (ordinary and IQHEc, paired and Hall metal states) in eight first subbands of the LL structure. These fillings are visualized in Fig. \ref{fig:1}, where various types of ordering are represented by means of  distinct height spikes corresponding to filling fraction each, at which some correlated state can be realized.  The evolution of FQHE with increasing number $n$ of the LL is the most visible property. The gradual pushing of FQHE toward the edges of bands in higher LLs is noticeable. The structure of fillings is the same for both spin subbands of each LL. In the center of each subband for $n\geq 1$, the paired state occurs associated by $2n$ 'satellite' states of IQHEc type. Such states do not exist in the LLL, though in the center of both subbands with  $n=0$ also are located collective states, but of Hall metal type, not paired.

\begin{table}[h!]
\centering
\begin{tabular}{|p{1.6cm}|p{5.6cm}|p{6.5cm}|p{3.5cm}|}
\hline
LL  subb.&IQHE, IQHEc, IQHE paired&FQHE ($q-odd,\; l=1,2,3,\dots)$
&Hall metal \\
\hline
$0\uparrow$& 1 & 
$\frac{1}{q}, 1-\frac{1}{q}, \frac{l}{l(q-1)\pm1},1-\frac{l}{l(q-1)\pm 1}$&$\frac{1}{q-1},
1-\frac{1}{q-1}$ \\
\hline
$0\downarrow$&2& $1+\frac{1}{q}$,
$2-\frac{1}{q}, 1+\frac{l}{l(q-1)\pm 1}, 2-\frac{l}{l(q-1)\pm 1}$& $1+\frac{1}{q-1},
2-\frac{1}{q-1} $
\\
\hline
$1\uparrow$&$\frac{7}{3},\frac{8}{3},3$,
($\frac{5}{2}\; paired$)& $2+\frac{1}{3q},2+\frac{l}{3l(q-1)\pm 1},3-\frac{1}{3q}, 3-\frac{l}{3l(q-1)\pm 1} $& $2+\frac{1}{3(q-1)},3-\frac{1}{3(q-1)}$ \\
\hline
$1\downarrow$&$\frac{10}{3},\frac{11}{3},4$,($\frac{7}{2}\; paired$) &$3+\frac{1}{3q}, 3+\frac{l}{3l(q-1)\pm 1}, 4-\frac{1}{3q},4-\frac{l}{3l(q-1)\pm1}$&
$3+\frac{1}{3(q-1)}, 4-\frac{1}{3(q-1)}$\\
\hline
$2\uparrow$&$\frac{21}{5},\frac{22}{5},\frac{23}{5},\frac{24}{5},5$,($\frac{9}{2}\; paired$) &
$4+\frac{1}{5q}, 4+\frac{l}{5l(q-1)\pm 1},5-\frac{1}{5q}, 5-\frac{l}{5l(q-1)\pm 1}$&$ 4+\frac{1}{5(q-1)}, 5-\frac{1}{5(q-1)}$\\
\hline
$2\downarrow$&$\frac{26}{5}, \frac{27}{5}, \frac{28}{5}, \frac{29}{5},6$,($\frac{11}{2}\; paired$) &
$5+\frac{1}{5q}, 5+\frac{l}{5l(q-1)\pm 1},
6-\frac{1}{5q}, 6-\frac{l}{5l(q-1)\pm 1}$&$5+\frac{1}{5(q-1)}, 6-\frac{1}{5(q-1)}$\\
\hline
$3\uparrow$&$\frac{43}{7}, \frac{44}{7}, \frac{45}{7}, \frac{46}{7}, \frac{47}{7}, \frac{48}{7}, 7$,($\frac{13}{2}\; paired$) &
$6+\frac{1}{7q}, 6+\frac{l}{7l(q-1)\pm 1},
7-\frac{1}{7q},7-\frac{l}{7l(q-1)\pm 1}$&$ 6+\frac{1}{7(q-1)}, 7-\frac{1}{7(q-1)}$\\
\hline
$3\downarrow$&$\frac{50}{7}, \frac{51}{7}, \frac{52}{7}, \frac{53}{7}, \frac{54}{7}, \frac{55}{7}, 8$,($\frac{15}{2}\; paired$) &
$7+\frac{1}{7q},
7+\frac{l}{7l(q-1)\pm 1},8-\frac{1}{7q},8-\frac{l}{7l(q-1)\pm 1}$&
$7+\frac{1}{7(q-1)}, 8-\frac{1}{7(q-1)}$\\
\hline
\end{tabular}
\caption{LL filling factors for IQHE and FQHE determined by commensurability arguments ($paired$ indicates condensate of electron pairs),  for $l=\frac{4}{3},\frac{5}{3},\frac{3}{2}\dots$ one can obtain in subband $0\uparrow$ all rates out of the main line, cf. note \cite{uwaga} and Tab. \ref{tab2} (note that $\pm $ in the denominators can be  formally substituted by $l \rightarrow \pm l $)}
\label{tab1} 
\end{table}

Special attention would be paid to the filling factors from the LLL which cannot be obtained by the formula (\ref{frac}) with $l$ integer. Examples of such fillings are $\frac{5}{13},\;\frac{4}{11}, \;\frac{4}{5},\;\frac{5}{7},\;\frac{3}{8},\;\frac{3}{10}$, for which FQHE is experimentally observed \cite{pan2003}. One can notice that if one assumes 
$l=\frac{4}{3}$, which corresponds formally to $\frac{3hc}{4e}$ fragment of the flux quantum per particle (the result of division of the total flux per number of particles), and this fraction would correspond to residual flux through the last loop of threelooped trajectory. Then, by virtue of the formula (\ref{frac}) we get for $q=3$ and $l=\frac{4}{3}$,
\begin{equation}
\nu=\frac{4/3}{4/3(3-1)\pm 1}=\left\{\begin{array}{l}
\frac{4}{11},\\
\frac{4}{5}.\\
\end{array}\right.
\end{equation}
Similarly, for $l=\frac{5}{3}$ (hole symmetric fraction to $\frac{4}{3}$) and $q=3$ we get,
\begin{equation}
\nu=\frac{5/3}{5/3(3-1)\pm 1}=\left\{\begin{array}{l}
\frac{5}{13},\\
\frac{5}{7}.\\
\end{array}\right.
\end{equation}

\subsection{Generalization of the formula (\ref{frac})} 
In the case of multilooped cyclotron orbits ($q$-looped) we deal with FQHE in the particular LL subband, $n\uparrow (\downarrow)$, with the main fractional Hall structure (cf. Tab. \ref{tab1}),
\begin{equation}
\label{calosc}
\nu=\left\{\begin{array}{l}
2n+\frac{1}{(2n+1)q},\; (2n+1)-\frac{1}{(2n+1)q},\;for\; \uparrow,\\
(2n+1)+\frac{1}{(2n+1)q},\; (2n+2)-\frac{1}{(2n+1)q},\;for\;\downarrow.\\
\end{array}\right.
\end{equation}
If now the filling fraction $\nu=\frac{N}{N_0}$ is different that one given by Eq. (\ref{calosc}), then 
the total flux of $B$ field per particle in this particular subband equals to,
\begin{equation}
\left\{ \begin{array}{l}
\frac{BS}{N-2n N_0},\; for\; \uparrow,\\
\frac{BS}{N-(2n+1)N_0},\; for\; \downarrow.\\
\end{array}\right. 
\end{equation} 
The cyclotron orbit (q-looped) in this subband has the same size for both $\uparrow$ and $\downarrow$,
\begin{equation}
\frac{(2n+1)hc}{eB}=\frac{(2n+1)S}{N_0}. 
\end{equation}
If one assumes that $q-1$ loops take a full flux 'quantum' portion each (in the subband this 'quantum' portion of flux equals to $\frac{(2n+1)hc}{e}=\frac{(2n+1)BS}{N_0}$), then
the residual flux piercing the last loop equals to,
\begin{equation}
\Phi=\left\{ \begin{array}{l}
\frac{BS}{N-2n N_0}-(q-1)\frac{(2n+1)BS}{N_0},\; for\; \uparrow,\\
\frac{BS}{N-(2n+1)N_0} -(q-1)\frac{(2n+1)BS}{N_0},\; for\; \downarrow.\\
\end{array}\right. 
\end{equation}
In the case when the residual flux $\Phi$ coincides with the flux per particle for some filling fraction $l$ (in particular, corresponding to  certain completely filled   LL), $\Phi_0=\frac{BS}{lN_0}=\pm \Phi$, one arrives at the condition,
\begin{equation}
\label{fr3}
\nu=\left\{ \begin{array}{l}
2n+\frac{l}{(2n+1)l (q-1)\pm 1},\;(2n+1)-\frac{l}{(2n+1)l(q-1)\pm 1},\;for\;\uparrow,\\
(2n+1)+\frac{l}{(2n+1)l (q-1)\pm 1},\;(2n+2)-\frac{l}{(2n+1)l(q-1)\pm 1},\;for\;\downarrow,\\
\end{array}\right.
\end{equation}
corresponding to commensurability of the last loop of cyclotron orbit with 
order arrangement for the fraction $l$ ($\pm$ corresponds to consonant or opposite orientation of the residual flux, i.e., to eight-figure-shape traversing of the last loop for opposite orientation). 
In the limit $l\rightarrow \infty$ one gets the condition for the Hall metal,
\begin{equation}
\label{fr4}
\nu=\left\{ \begin{array}{l}
2n+\frac{1}{(2n+1) (q-1)},\;(2n+1)-\frac{1}{(2n+1)(q-1)},\;for\;\uparrow,\\
(2n+1)+\frac{1}{(2n+1) (q-1)},\;(2n+2)-\frac{1}{(2n+1)(q-1)},\;for\;\downarrow,\\
\end{array}\right.
\end{equation}
when for the last loop falls the zero flux ($\lim_{l\rightarrow \infty}\frac{BS}{lN_0}=0$) resulting in field-free motion as in the Fermi liquid without the magnetic field presence \cite{vor3,halperin,vor7}. 

\begin{figure}[ht]
\centering
\scalebox{1.6}{\includegraphics{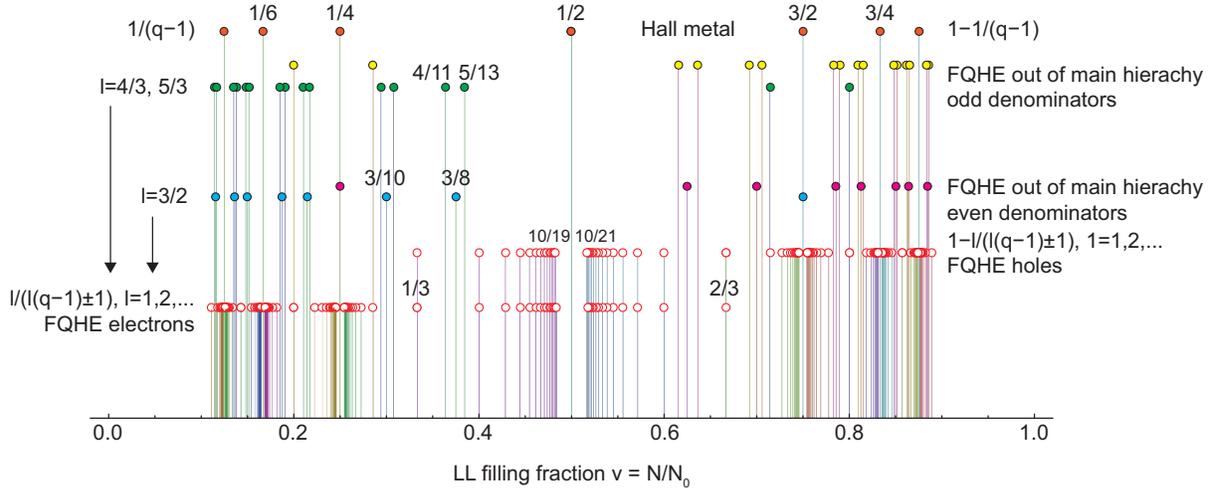}}
\caption{Graphical presentation of filling fractions in the  LLL subband $0\uparrow$ selected by commensurability condition, as listed in Table \ref{tab1}; filling fraction $\nu=\frac{N}{N_0}=\frac{B_0}{B}$, $B_0$ is magnetic field corresponding to $\nu=1$, by spikes of various height different types of ordering are indicated; only few FQHE series are displayed: $l=1-15$ with $q=3,5,7,9$ in Eq. (\ref{fr1}), moreover, series for $l=\frac{4}{3},\frac{5}{3},\frac{3}{2}$ with $q=3,5,7,9$ in Eq. (\ref{fr1}), main fractions acc. to Eq. (\ref{fr0}) for $q=3,5,7,9$, Hall metal states (limit $l \rightarrow \infty$ in Eq. (\ref{fr1})) for $q=3,5,7,9$; a few selected ratios from these series  are explicitly  written   }
\label{fig500}
\end{figure}

\begin{figure}[ht]
\centering
\scalebox{1.5}{\includegraphics{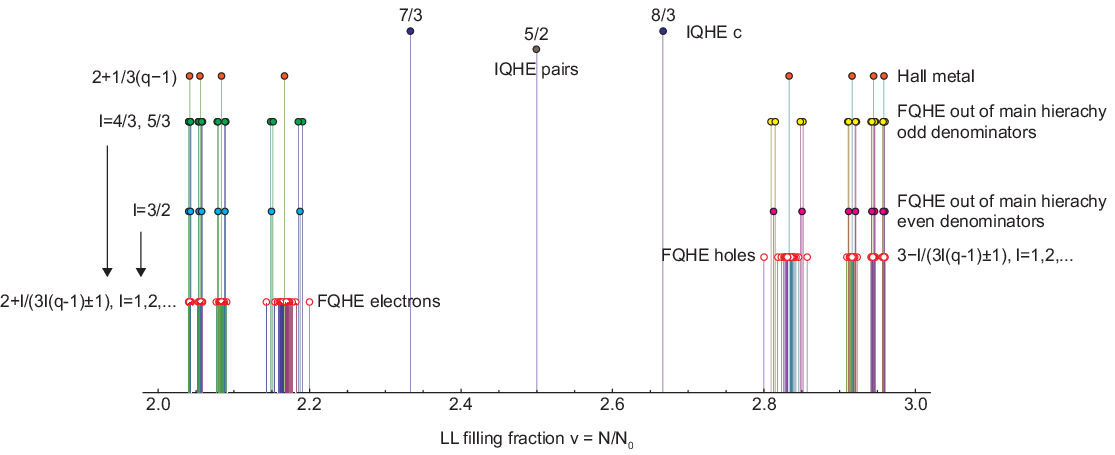}}
\caption{Graphical presentation of filling fractions in the $n=1$ LL subband $1\uparrow$ selected by commensurability condition, as listed in Table \ref{tab1}; filling fraction $\nu=\frac{N}{N_0}=\frac{B_0}{B}$, $B_0$ is magnetic field corresponding to $\nu=1$; only few FQHE series are displayed---the same as in Fig. \ref{fig500} for $0\uparrow$ subband, i.e., $l=1-15$ with $q=3,5,7,9$ in Eq. (\ref{fr3}) for $n=1$, moreover, series for $l=\frac{4}{3},\frac{5}{3},\frac{3}{2}$ with $q=3,5,7,9$ in Eq. (\ref{fr3}), main fractions acc. Eq. (\ref{calosc}) for $q=3,5,7,9$, Hall metal states (limit $l \rightarrow \infty$ in Eq. (\ref{fr3}) and $n=1$) for $q=3,5,7,9$; a few selected ratios from these series  are explicitly  written; new features for $n\geq 1$, i.e., $\frac{7}{3},\frac{8}{3}$ in the central part of the subband correspond to single-looped IQHE-type (indicated as IQHEc), at $\frac{5}{2}$ the correlated IQH-type state of pairs  }
\label{fig501}
\end{figure}

\begin{figure}[ht]
\centering
\scalebox{1.4}{\includegraphics{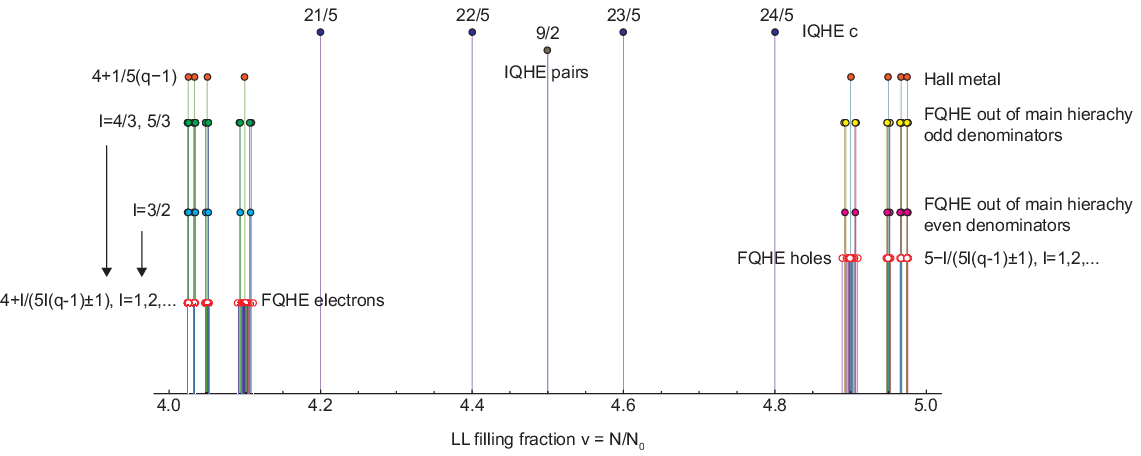}}
\caption{Graphical presentation of filling fractions in the $n=2$ LL subband $2\uparrow$ selected by commensurability condition, as listed in Table \ref{tab1}; filling fraction $\nu=\frac{N}{N_0}=\frac{B_0}{B}$, $B_0$ is magnetic field corresponding to $\nu=1$; only few FQHE series are displayed---the same as in Fig. \ref{fig500} and Fig. \ref{fig501} for subband $0\uparrow$ and $1\uparrow$, i.e., $l=1-15$ with $q=3,5,7,9$ in 
Eq. (\ref{fr3}) for $n=2$, moreover, series for $l=\frac{4}{3},\frac{5}{3},\frac{3}{2}$ with $q=3,5,7,9$ in Eq. (\ref{fr3}), main fractions acc. Eq. (\ref{calosc}) for $q=3,5,7,9$, Hall metal states (limit $l \rightarrow \infty$ in Eq. (\ref{fr3}) and $n=2$) for $q=3,5,7,9$; a few selected ratios from these series  are explicitly  written; new features twice in number in comparison to $n=1$ case, i.e., $\frac{21}{5},\frac{22}{5}, \frac{23}{5}, \frac{24}{5}$ in central part of the subband correspond to single-looped IQHE-type (indicated as IQHEc), at $\frac{9}{2}$ the correlated IQH-type state of pairs  }
\label{fig502}
\end{figure}

\begin{figure}[ht]
\centering
\scalebox{1.5}{\includegraphics{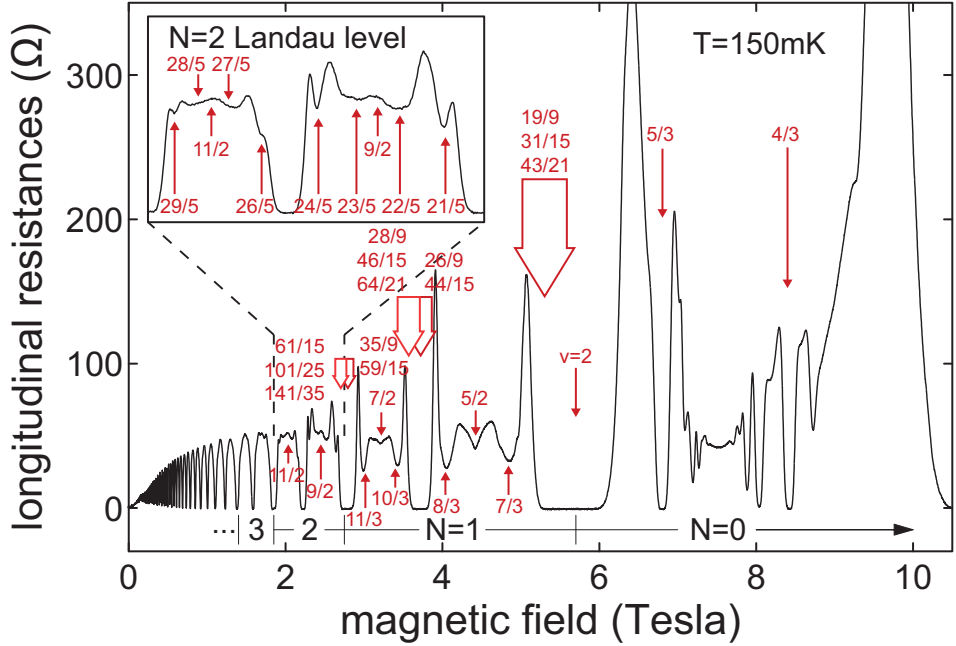}}
\caption{Resistivity measurements for wide range of magnetic field corresponding to $n=1,2$ in high mobility GaAs/AlGaAs heterostructure (after Ref. \cite{lls}), in color are indicated fractions predicted by commensurability arguments in perfect agreement with observed features---cf. Table \ref{tab1} and Fig. \ref{fig:1}}
\label{fig:9}
\end{figure}

\begin{table}[h!]
\centering
\begin{tabular}{|p{3.5cm}|p{6cm}|}
\hline
LL subband& LL filling ratios for Hall metal state \\
\hline
$0\uparrow $ & 1/2, 1/4, 1/6, 1/8, 1/10, 3/4, 5/6, 7/8, 9/10\\
\hline
$0\downarrow$&3/2, 5/4, 7/8, 9/8, 11/10, 7/4, 11/6, 15/8, 19/10\\
\hline
$1\uparrow$& 13/6, 25/12, 37/18, 49/24, 61/30, 17/6, 35/12, 53/18, 71/24, 89/30\\
\hline
$1\downarrow$&19/6, 37/12, 55/18, 73/24, 91/30, 23/6, 47/12, 71/18, 95/24, 119/30\\
\hline
$2\uparrow$ &41/10, 81/20, 121/30, 161/40, 201/50, 49/10, 99/20, 149/30, 199/40, 249/50\\
\hline
$2\downarrow$&51/10, 101/20, 151/30, 201/40, 251/50, 59/10, 119/20, 179/30, 239/40, 299/50\\
\hline
\end{tabular}
\caption{The LL filling factors for possible Hall metal states determined by the commensurability condition, i.e., by Eq. (\ref{fr4})---the limit $l\rightarrow \infty$ of the Eq. (\ref{fr3}), for $q=3,5,7,9,11$}
\label{tab3}
\end{table}

\begin{figure}[ht]
\centering
\scalebox{1.5}{\includegraphics{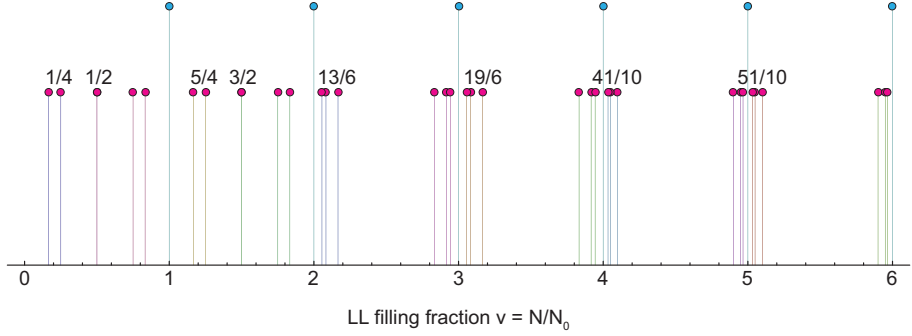}}
\caption{Graphical presentation of evolution of Hall metal type states with growing number of LL subband; selected by commensurability condition, as listed in Tab. \ref{tab3} and given by Eq. (\ref{fr4}); filling fraction $\nu=\frac{N}{N_0}=\frac{B_0}{B}$, $B_0$ is magnetic field corresponding to $\nu=1$; a few  fractions are explicitly written, for $n>0$---the most centrally located }
\label{fig505}
\end{figure}

\begin{figure}[ht]
\centering
\scalebox{1.5}{\includegraphics{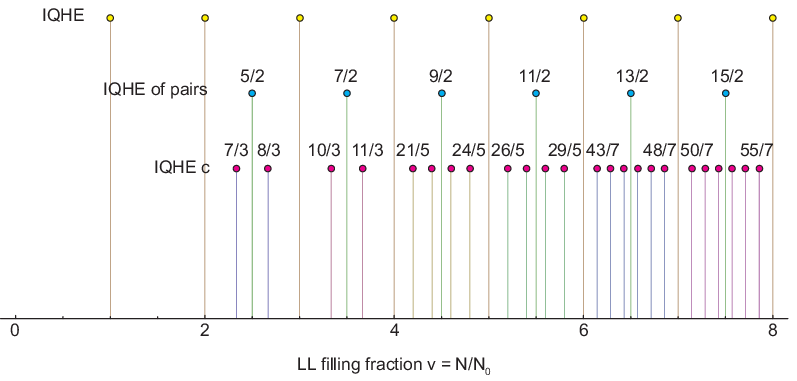}}
\caption{Graphical presentation of new Hall features in higher LLs occurring exclusively for $n\geq 1$ ($\nu>2$) identified  by the commensurability condition, as listed in Tab. \ref{tab1} (cf. Fig. \ref{fig:1}); filling fraction $\nu=\frac{N}{N_0}=\frac{B_0}{B}$, $B_0$ is magnetic field corresponding to $\nu=1$; some fractions are explicitly written for IQHEc states; at $\nu=\frac{5}{2},\frac{9}{2},\frac{11}{2},\frac{13}{2},\frac{15}{2}$ the correlated states are of the paired type }
\label{fig510}
\end{figure}

In Figs \ref{fig500}, \ref{fig501} and \ref{fig502} the comparison of fractional filling structure for quantum Hall features is presented for subbands $0\uparrow$, $1\uparrow$, $2\uparrow$, respectively. In these three schemes the same fraction families are indicated and the evolution of the filling hierarchy with increasing LL $n$ number is apparent. In Fig. \ref{fig505} the filling fractions for the  Hall metal state are shown for few first LLs according to Tab. \ref{tab3}. The evolution of new features in LLs with $\geq 1$ is visualized in Fig. \ref{fig510}.    

\subsection{Fractions with even denominators and pairing} 

The commensurability conditions allow also for systematic analysis of some special fractional fillings of LLs expressed by ratios with even denominators. The most prominent seem to be the ratios exclusively in $n>0$ LLs of the form, $\nu =\frac{5}{2}, \frac{7}{2}, \frac{9}{2}, \frac{11}{2} \dots$, resulting from the special commensurability condition,
 \begin{equation}
\label{frac3}
\frac{(2n+1)hc}{eB}=\frac{\frac{2n+1}{2}S}{N-\left\{
\begin{array}{ll}
2n N_0 & for\;\uparrow, \\
(2n+1)N_0 & for \;\downarrow,
\end{array}\right.}
\end{equation}
 for subbands $n\uparrow (\downarrow)$, respectively. By pairing of particles in the last subband the ideal commensurability can be achieved for the pairs allowing IQH ordering for them (this is due to twice reducing of denominator in r.h.s. of the above formula for pairs, while the cyclotron orbits conserve their size for pairs because cyclotron radius scales as $\sim\frac{e}{m}=\frac{2e}{2m}$). This happens for $\frac{5}{2},\frac{7}{2},\frac{9}{2}, \dots$, but not for $\nu=\frac{1}{2}$ and $\frac{3}{2}$ from $n=0$ spin subbands, where cyclotron orbits always are shorter than separation of particles, what precludes pairing. The latter two ratios correspond to Hall metal states, are given by the conditions which can be obtained from formulae (\ref{fr1}) and (\ref{fr2}) taken in the limit $l\rightarrow\infty$. These features for $n=0,1,2$ remarkably agree with experimental observations and numerical analysis of states proposed for related filling ratios as it has been summarized recently in Ref. \cite{5/2-1}.

\subsection{Multicomponent wave functions for higher LL subband fractional fillings}
One can use the multicomponent model of FQHE in the form of the Halperin wave function \cite{halperin1983,goerbig2007} in order to account for variously correlated fractions of the total system. The Halperin wave function was a generalization to SU(2) of the Laughlin function (with the magnetic length, $l_m=1$),
\begin{equation}
\Psi^L_m(\{z_k\})=\prod_{k<l}^N(z_k-z_l)^m e^{-\sum_k^N|z_k|^2/4},
\end{equation}
to two-spin component system in the following form,
\begin{equation}
\Psi^L_{m\downarrow}(\{z_{k\downarrow}^{\downarrow}\}) \Psi^L_{m\uparrow}(\{z_{k\uparrow}^{\uparrow}\}), 
\end{equation}
with possible inter-component correlation factor, 
\begin{equation}
\prod_{k\downarrow}^{N\downarrow}\prod_{k\uparrow}^{N\uparrow}(z_{k\downarrow}^{\downarrow}
-z_{k\uparrow}^{\uparrow})^n.
\end{equation}
where the exponent $n$ may be even or odd, while $m$, $m\downarrow$ and $m\uparrow$ are odd.
The two component Halperin function \index{Halperin function} can be easy generalized to SU(N) case \cite{goerbig2007}, and in the case of graphene with the spin-valley SU(4) symmetry was introduced in the form \cite{papic2009a},
\begin{equation}
\begin{array}{ll}
\Psi^{SU(4)}_{m_1,...,m_4,n_{i,j}}&=\prod_j^4\prod_{k_j<l_j}^{N_j}(z_{k_j}^j-z_{l_j}^j)^{m_j}e^{-\sum_{j=1}^4\sum_{k_j=1}^{N_j}
|z_{k_j}^j|^2/4}\\
&\times \prod_{i<j}^4\prod_{k_i}^{N_i}\prod_{k_j}^{N_j}(z_{k_i}^i-z_{k_j}^j)^{n_{i,j}},\\
\end{array}
\end{equation}
$m_j$ must be odd integers, whereas $n_{ij}$ may be also even integers.
Since the components of this function have Jastrow polynomial \index{Jastrow polynomial} form, the similar topological interpretation is required as in the single component electron liquid case. 
Such type of wave functions might be used to model trial wave functions corresponding to FQHE/IQHEc with appropriate exponents related to particular fractions of electrons selected by commensurability conditions in the last subband including IQH ordering in the rest of subbands completely filled.

\section{Comparison with experiment}

\begin{figure}[th]
\centering
\scalebox{1.2}{\includegraphics{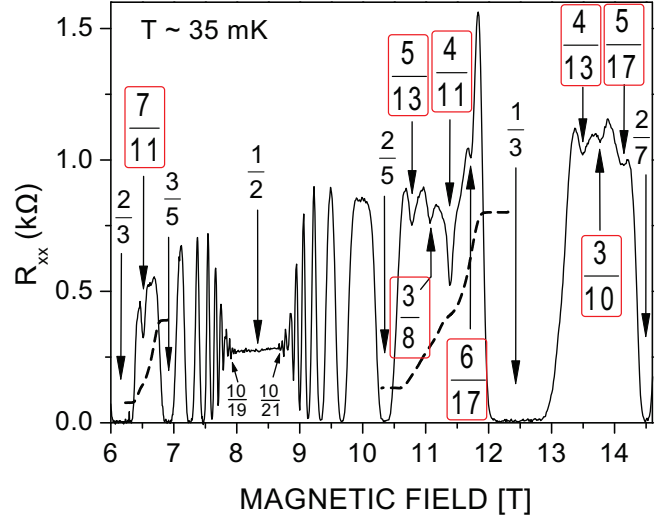}}
\caption{FQHE features in a quantum well GaAs/AlGaAs with electron density of $10^{11}$ 1/cm$^{-2}$; $R_{xx}$ for $\frac{2}{3} > \nu > \frac{2}{7}$ at the temperature equal $T \sim 35$ mK; Hall resistance $R_{xy}$ in the region of $\nu = \frac{7}{11}$ and $\nu = \frac{4}{11}$ marked with the dotted line (after Ref. \cite{pan2003}); all ratios have been successfully explained by commensurability condition---cf. Tab. \ref{tab2}, fractions outside  the  line of hierarchy (\ref{frac}) with $l=1,2,\dots$, are indicated in red }
\label{fig:55}
\end{figure}

The LL filling structure summarized in Tab. \ref{tab1} and in Fig. \ref{fig:1} reveals the asymmetry of FQHE/IQHEc organization in the higher LLs in comparison to the lowest one in agreement with the experimental observations \cite{pan2003,lls}. The hierarchy of filling ratios determined by presented above topological conditions very well reproduces the FQHE manifestation in the LLL, in its two spin bands, $0\uparrow$ and $0\downarrow$ in close agreement with the magneto-resistance measurements in semiconductor 2DEG as presented in Fig. \ref{fig:55} with the data after Ref. \cite{pan2003} (cf. Tab. \ref{tab2}). Besides the main line hierarchy as given by formula (\ref{fr1}) (the same as in the standard CF theory), the ratios outside the CF predictions were also reproduced, like $\nu=\frac{4}{11},\frac{5}{13}, \frac{4}{5},\frac{5}{7},\frac{3}{8},\frac{3}{10}$---they are indicated in Fig. \ref{fig:55}. The LLL subband filling rates  which are  possible to be predicted by commensurability condition (for electrons) are collected in Tab. \ref{tab2}. The completely different of FQHE/IQHEc hierarchy in the $n=1,2$ LLs is determined by topology arguments and it finds a pretty good coincidence with the data presented in Ref. \cite{lls} as it has been indicated in Fig. \ref{fig:9}. The natural explanation of difference between the LLL and the LLs with $n=1,2$ with respect to FQHE manifestation is the convincing success of topological approach especially in view of different predictions of standard CF theory failed in face of the experimental data.

\begin{figure}[th]
\centering
\includegraphics{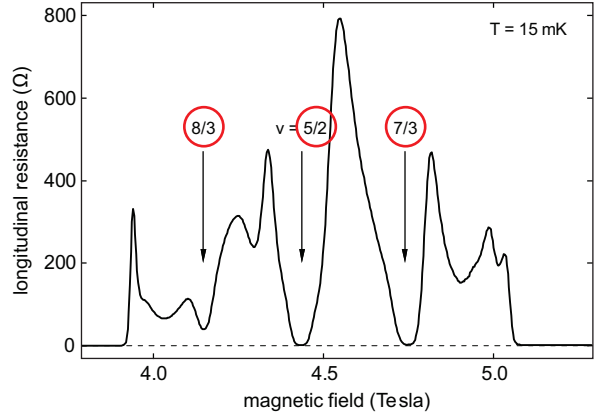}
\caption{Experimentally observed resistivity minima in center of of the subband $1 \uparrow $ of the $n=1$ LL (after Ref. \cite{lls}), filling ratios predicted by commensurabilty arguments are marked }
\label{fig:8}
\end{figure}

To be specific, let us note that in $1\uparrow$ LL subband the minima of longitudinal resistivity were observed at $\nu=\frac{7}{3}$ and $\frac{8}{3}$ accompanying also the minimum at $\nu=\frac{5}{2}$---cf. Fig.\ref{fig:8}. They were interpreted as manifestation of FQHE in this subband. Such modest FQHE features in $1\uparrow$ subband are in contrast with the abundance of FQHE in the lowest LL. Both fractions $\frac{7}{3}$ and $\frac{8}{3}$ can be predicted based on commensurability condition, though from this condition it follows that these states would be of IQHEc type with singlelooped cyclotron orbits. The state at $\frac{5}{2}$ ought be paired.

\begin{figure}[th]
\centering
\scalebox{0.68}{\includegraphics{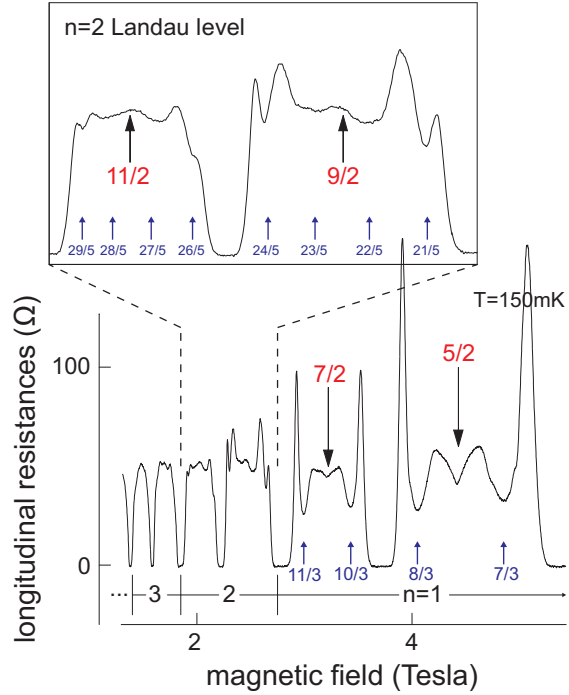}}
\caption{Resistivity measurements for magnetic field corresponding to $n=1,2$ LLs in high mobility ($11\times 10^6$ cm$^2$/Vs) GaAs/AlGaAs heterostructure (after Ref. \cite{lls}), with blue arrows are indicated fractions predicted by commensurability arguments as IQHEc, in red are marked fractions with pairing predicted by commensurabilty---Tab. \ref{tab1} and Fig. \ref{fig:1}}
\label{fig:19}
\end{figure}

The similar situation is repeated in the $1\downarrow$ subband where the doublet of minima 
$\frac{10}{3}$ and $\frac{11}{3}$ accompanies the minimum $\frac{7}{2}$. These correlated states have been also successfully predicted by commensurability arguments (cf. Tab. \ref{tab1}). 
In the next subbands $2\uparrow$ and $2\downarrow$ which belong to $n=2$ LL, the
situation again changes and two states with even denominator $\frac{9}{2}$ and $\frac{11}{2}$ (paired) are accompanied by four states with IQHEc ordering, $\frac{21}{5},\frac{22}{5},\frac{23}{5}, \frac{24}{5}$ and $\frac{26}{5}, \frac{27}{5}, \frac{28}{5}, \frac{29}{5}$, respectively. These predicted features are in excellent coincidence with the experimental observations \cite{lls} as is illustrated in Fig. \ref{fig:9} and in Fig. \ref{fig:19}. 
The experimentally observed features quite well correspond to IQHEc states as listed in Tab. \ref{tab1}. 

Let us note, that the states $\frac{8}{3},\frac{7}{3}$ have been investigated both theoretically and experimentally \cite{ll8/3-1,ll8/3-2}. These states are linked to $\frac{5}{2}$ state regarded as FQHE for filling fraction with even denominator \cite{5/2-1}, Fig. \ref{fig:8}. The fractional type of Hall effect is suggested for these states and possibility of non-Abelian charges convenient for topological quantum information processing is considered \cite{5/2-1,ll8/3-2,ll8/3-1}. Hence, the suggestion that states like $\frac{8}{3} , \frac{7}{3} $ are not of FQHE type but rather of IQHEc type seems to contribute importantly to this discussion. 

The described above states of IQHEc type constitute the most characteristic feature of higher LLs fractional hierarchy. The number of these states in particular subband increases with the number $n$ of consecutive LLs---for $n=1$ there are two such states, for $n=2$---four, while for $n=3$---six, cf. Tab. \ref{tab1}, in general, as $2n$. Besides  these features in higher LLs the another characteristic property is also predicted by commensurability conditions, i.e., the true FQHE states in the higher Landau LLs (for $n\geq 1$) characterized by multilooped cyclotron structure, similarly as in the LLL (with $n=0$). Due to commensurability constraints these true FQH states are pushed toward the subband edges, the closer to the edges the higher LL number is. This property is clearly visualized in Fig. \ref{fig:1}. To match these prediction with experimental observations we note that the series of FQH states in the higher LLs are densely located and concentrated in close vicinity of subband rims (cf. Tab. \ref{tab1} and Fig. \ref{fig:1})  and may be beyond the resolution ability of measurement technique. Closely laying FQH states in vicinity of IQH state may be not distinguishable with the applied resolution of observation and thus would mix together resulting in larger flattening of main IQHE minimum what is actually observed \cite{lls}. In Fig. \ref{fig:9}  this situation is marked by thick arrows.

\section{Conclusion}
The topological braid group-based approach to FQHE/IQHE turns out to be effective in recognition of fractional fillings in the higher LLs and in explanation of distinct structure of FQHE in these levels in comparison to the LLL case. The riches of fractional fillings of the LLL at which FQHE manifests itself stays in striking dissimilarity in the higher LLs, where the structure of fillings from the LLL is not repeated and FQHE is not so frequent there. The reason of this property is in distinct commensurability of cyclotron orbits and interparticle separation in various LL subbands. With growing number of the occupied LL, the corresponding cyclotron orbit becomes larger due to growing kinetic energy and thus the exceeding of its size by interparticle distances, critical for FQHE, would happen in higher subbands only at relatively small density of particles. Thus FQHE is gradually pushed toward the edges of the subband with growing subband number in opposition to the LLL, inside which cyclotron orbits were always too short for particle interchanges. Simultaneously, in the higher LLs some new commensurability opportunities occur which were impossible in the LLL. This new property resembles commensurability in completely filled subbands thus corresponds to singlelooped orbits characteristic for IQHE, but without the large energy gap protecting this special filling as for completely filled LL subbands. In the LLs with $n>0$ the increasing  number, $2n$, of these IQHEc states symmetrically positioned  with respect to in middle  located paired state, occupy central part of higher LL subbands emptying of FQH states, which are gradually pushed toward subband edges. The filling ratios selected by this type of commensurability are visible in experiment similarly as also predicted paired states in centers of subbands for $n\geq 1$. 
The explanation ability for structure of fractional fillings of higher LLs together with direct experimental evidence for triggering role of carrier mobility in FQHE formation \cite{jac2013} support usefulness of cyclotron braid group approach to collective states in 2D charged system in the presence of quantizing magnetic field.

\begin{acknowledgments}
Authors acknowledge the support of the present work upon the NCN project no. 2011/02/A/ST3/00116.
\end{acknowledgments}


\begin{thebibliography}{10}
\bibitem{uwaga}
for '-' in the denominator in Eq. (\ref{frac}), $l$=2\dots; moreover, for
  not-integer $l$, but for some characteristic fillings above the first band,
  e.g., $l=\frac{4}{3},\frac{5}{3},\frac{3}{2}$, where $l$ displays in
  (\ref{frac}) the filling fraction accommodated to the last loop of multiloop
  cyclotron trajectory, one obtains acc. to the concept of commensurability,
  the fractions $\nu$ out of the main line,
  $\nu=\frac{4}{11},\frac{4}{5},\frac{5}{13},\frac{5}{7},\frac{3}{8},\frac{3}{10}\dots$.
\end{thebibliography}
\end{document}